\documentclass{aa}  
\usepackage{amsmath}
\usepackage{comment}
\usepackage{graphicx}
\usepackage{txfonts}
\usepackage{color}
\usepackage[colorlinks]{hyperref}

\hypersetup{
    colorlinks = true,
    linkcolor = {blue},
    citecolor = {blue},
}

\usepackage{array}
\usepackage{multirow}
\newcolumntype{P}[1]{>{\centering\arraybackslash}p{#1}}
\newcolumntype{D}{ >{\centering\arraybackslash} c{1cm} }
\usepackage[table, dvipsnames]{xcolor}

\definecolor{AS}{RGB}{201, 123, 132}

\definecolor{bg_visc}{RGB}{249, 235, 185}
\definecolor{bg_visc_light}{RGB}{255, 254, 234}

\definecolor{bg_mhd}{RGB}{252, 202, 214}
\definecolor{bg_mhd_light}{RGB}{255, 242, 242}

\definecolor{bg_2}{RGB}{218, 229, 215}

\defcitealias{Somigliana+2022}{S22}

\begin{document}

   \title{The evolution of the \texorpdfstring{$M_{\mathrm{d}} - M_{\star}$}{Mdisc-Mstar} and \texorpdfstring{$\dot M - M_{\star}$}{Mdot-Mstar} correlations traces protoplanetary disc dispersal}


   \author{Alice Somigliana
          \inst{1,2}\thanks{email: alice.somigliana@eso.org},
          Leonardo Testi\inst{3,4},
          Giovanni Rosotti\inst{5},
          Claudia Toci\inst{1},
          Giuseppe Lodato\inst{5},
          Rossella Anania\inst{5},
          Benoît Tabone\inst{6},
          Marco Tazzari\inst{4},
          Ralf Klessen\inst{7, 8},
          Ugo Lebreuilly \inst{9},
          Patrick Hennebelle \inst{9},
          and Sergo Molinari \inst{10}
          }

   \institute{European Southern Observatory, Karl-Schwarzschild-Strasse 2, D-85748 Garching bei München, Germany\\
              \email{alice.somigliana@eso.org}
         \and
             Fakultat für Physik, Ludwig-Maximilians-Universität München, Scheinersts. 1, 81679 München, Germany
         \and
             Dipartimento di Fisica e Astronomia, Universita' di Bologna, Via Gobetti 93/2, I-40122 Bologna, Italy
         \and
             INAF-Osservatorio Astrofisico di Arcetri, Largo E. Fermi 5, I-50125 Firenze, Italy
        \and 
             Dipartimento di Fisica, Università degli Studi di Milano, Via Celoria 16, I-20133 Milano, Italy
        \and
             Université Paris-Saclay, CNRS, Institut d'Astrophysique Spatiale, Orsay, France
        \and
             Universität Heidelberg, Zentrum für Astronomie, Institut für Theoretische Astrophysik, Albert-Ueberle-Str. 2, D-69120 Heidelberg, Germany
        \and
             Universität Heidelberg, Interdisziplinäres Zentrum für Wissenschaftliches Rechnen, Im Neuenheimer Feld 205, D-69120 Heidelberg, Germany
        \and
             Université Paris-Saclay, Université Paris Cité, CEA, CNRS, AIM, 91191, Gif-sur-Yvette, France
        \and
            Istituto Nazionale di Astrofisica-IAPS, Via Fosso del Cavaliere 100, I-00133 Roma, Italy
             }

   \date{Received XXX; accepted YYY}

 
  \abstract{
  Observational surveys of entire star-forming regions have provided evidence of power-law correlations between the disc integrated properties and the stellar mass, especially the disc mass ($M_{\mathrm{d}} \propto {M_{\star}}^{\lambda_{\mathrm{m}}}$) and the accretion rate ($\dot M \propto {M_{\star}}^{\lambda_{\mathrm{acc}}}$). Whether the secular disc evolution affects said correlations is still a matter of debate: while the purely viscous scenario has been investigated, other evolutionary mechanisms could have a different impact. In this paper, we study the time evolution of the slopes $\lambda_{\mathrm{m}}$ and $\lambda_{\mathrm{acc}}$ in the wind-driven and viscous-wind hybrid case and compare it to the purely viscous prediction. We use a combination of analytical calculations, where possible, and numerical simulations performed with the 1D population synthesis code \texttt{Diskpop}, that we also present and release to the community. Assuming $M_{\mathrm{d}}(0) \propto {M_{\star}}^{\lambda_{\mathrm{m}, 0}}$ and $\dot M(0) \propto {M_{\star}}^{\lambda_{\mathrm{acc}, 0}}$ as initial conditions, we find that viscous and hybrid accretion preserve the power-law shape of the correlations, while evolving their slope; on the other hand, MHD winds change the shape of the correlations, bending them in the higher or lower end of the stellar mass spectrum depending on the scaling of the accretion timescale with the stellar mass. However, we show how a spread in the initial conditions conceals this behaviour, leading to power-law correlations with evolving slopes like in the viscous and hybrid case. We analyse the impact of disc dispersal, intrinsic in the wind model and due to internal photoevaporation in the viscous case: we find that the currently available sample sizes ($\sim 30$ discs at 5 Myr) introduce stochastic oscillations in the slopes evolution, which dominate over the physical signatures. We show that we could mitigate this issue by increasing the sample size: with $\sim 140$ discs at 5 Myr, corresponding to the complete Upper Sco sample, we would obtain small enough error bars to use the evolution of the slopes as a proxy for the driving mechanism of disc evolution. Finally, from our theoretical arguments we discuss how the observational claim of steepening slopes necessarily leads to an initially steeper $M_{\mathrm{d}} - M_{\star}$ correlation with respect to $\dot M - M_{\star}$.

  }


   \titlerunning{The $M_{\mathrm{d}}-M_{\star}$ and $\dot M - M_{\star}$ correlations trace disc dispersal}
   \authorrunning{Somigliana, A. et al.}
    
   \maketitle

%

\section{Introduction}

The secular evolution of protoplanetary discs is deeply intertwined with both the planet formation process \citep{Morbidelli2012-planetformreview} and the accretion onto the central protostar \citep{Hartmann1998}. Planetesimals, the building blocks of planets, form and evolve within the disc following the dynamics of either the gaseous or solid component, depending on their relative size and their coupling (or lack thereof) with the gas particles; on the other hand, the protostar is fed by the disc itself, through the accretion of material that loses angular momentum and drifts inwards. The ideal ground to explore the connection between protoplanetary discs and their host stars is provided by large surveys of entire star-forming regions, targeting the properties of both discs and protostars; the last decade has seen a significant observational effort in the direction of these population-level studies, also thanks to the advent of facilities like the Atacama Large Millimeter Array (ALMA) (see the PPVII reviews by \citealt{Manara+2023PPVII, Miotello+2023PPVII}). 

Disc masses and accretion rates are arguably the most studied integrated disc properties. Accretion rates are inferred from the spectra of the central stars, which show an excess emission (especially prominent in the UV) when accretion is taking place; surveys performed across different star-forming regions \citep{Muzerolle+2003, Natta+2004-lowmassaccrates, Mohanty+2005, DullemondNattaTesti2006, Herczeg&Hillenbrand2008, Rigliaco+2011, Manara+2012, Alcala+2014, Manara+2016-ChaI, Alcala+2017, Manara2017-Cha, Venuti+2019accrates, Manara+2020-UpperSco} agree on the presence of a power-law correlation between the accretion rate and the stellar mass, $\dot M \propto {M_{\star}}^{\lambda_{\mathrm{acc, obs}}}$ (following the notation of \citealt{Somigliana+2022}). On the other hand, disc masses have traditionally been determined from observations of the sub-mm continuum emission of the solid component of discs; due to the large number of assumptions involved in converting sub-mm fluxes into total disc masses (see \citealt{Miotello+2023PPVII}), one of the current main goals of the protoplanetary disc community is the accurate determination of total disc masses - both from dynamical constraints \citep{Veronesi+2021, Lodato+2023} and direct measurements of the total gas content (e.g., \citealt{Bergin+2013HD, Anderson+2022, Trapman+2022}). Despite the systematic uncertainties involved in their determination, dust-based disc masses also seem to show a power-law correlation with the stellar mass, $M_{\mathrm{d}} \propto {M_{\star}}^{\lambda_{\mathrm{m, obs}}}$, across different star-forming regions \citep{Ansdell+2016Survey, Ansdell+2017, Barenfeld2016-masssurvey, Pascucci+2016, Testi+2016, Testi+2022-Ofiucone, Sanchis+2020Lupus}.

The existence of the disc mass-stellar mass and accretion rate-stellar mass correlations is now generally accepted; however, there is no consensus on the physical reason behind their establishment and their evolution with time. While the $\dot M - M_{\star}$ correlation appears to have a roughly constant slope\footnote{Throughout this work, we use 'slope' as a synonym of power-law index, referring to the correlations in the logarithmic plane.} of $\lambda_{\mathrm{acc, obs}} \approx 1.8 \pm 0.2$ (as first suggested by \citealt{Muzerolle+2003} and supported by many of the following works mentioned above), the $M_{\mathrm{d}} - M_{\star}$ correlations is claimed to be steepening with time \citep{Ansdell+2017}, from the lowest $\lambda_{\mathrm{m, obs}} ( t \sim 1$ Myr$ ) = 1.7 \pm 0.2$ (Taurus) to the highest $\lambda_{\mathrm{m, obs}} (t \sim 5$ Myr$) = 2.4 \pm 0.4$ (Upper Sco). Whether these correlations reflect the initial conditions of disc populations, or are rather a product of the secular evolution, is still under debate. Both possibilities have been discussed for the $\dot M - M_{\star}$ correlation: \cite{AlexanderArmitage2006} have assumed it to hold as initial condition, favouring the correlation to be present in young populations, whereas \cite{DullemondNattaTesti2006} have derived it from a simple model of disc formation from a rotating collapsing core, which provided an explanation for evolved disc populations. At the same time, the claimed increase in the slope of $M_{\mathrm{d}} - M_{\star}$ does suggest an evolutionary trend; \cite{Somigliana+2022} have found that, assuming power-law correlations between both $M_{\mathrm{d}}$ and $\dot M$ and the stellar mass as initial conditions, secular evolution can indeed alter the slopes of the correlations themselves (see Section \ref{sec:time_evo_analytical} for details). However, their analysis was limited to the standard viscous evolution paradigm, whereas the driving mechanism of accretion is far from being constrained (see \citealt{Manara+2023PPVII} for a review). 

The traditional viscous accretion model prescribes a macroscopic viscosity as the cause of redistribution of angular momentum within the disc \citep{LyndenBellPringle1974, Pringle1981}. In this scenario, while part of the material loses angular momentum and moves radially closer to the star, some other material gains the same amount of angular momentum and moves further away, effectively increasing the disc size. The viscous paradigm can explain many key features of disc evolution, but it cannot account for disc dispersal - as determined from the observational evidence of exponentially decreasing fraction of both disc-bearing \citep{Hernandez+2007} and accreting \citep{Fedele+2010} sources in star-forming regions with time; furthermore, the low levels of turbulence detected in discs \citep{Pinte+2016, Flaherty+2018, Rosotti2023-Review} appear incompatible with the observed evolution. While the discrepancy in the disc and accretion fraction can be mended considering mechanisms such as internal or external photoevaporation \citep{Alexander+2014-PPVI, Winter+2018}, that effectively clear discs on timescales comparable with the observed decline, the tension between the expected and observed amount of turbulence does not appear to be solved yet. On the other hand, the MHD disc winds scenario offers a promising alternative. Pioneering work \citep{BlandfordAndPayne1982, Ferreira1997} supported by recent numerical simulations (e.g., \citealt{Bethune+2017}) demonstrated that MHD winds launched from the disc surface have the net effect of removing angular momentum as a consequence of the extraction of material; SMHD wind-driven accretion can even lead to disc dispersal \citep{ArmitageSimon&Martin2013, Tabone+2022LupusLetter}. Following disc evolution at population level in numerical simulations remains out of reach; however, three-dimensional core-collapse simulations have shown how non-ideal magnetohydrodynamics and ambipolar diffusion play a fundamental role in shaping the resulting population of early-type young stellar objects \citep{Lebreuilly+2021, Lebreuilly+2024}. While some 3D studies of isolated disc formation have attempted to bridge the gap between Class 0/I and Class II stages \citep{Machida&Hosokawa2013, Hennebelle+2020, Xu&Kunz2021a, Xu&Kunz2021b, Machida&Basu2024, Mauxion+2024}, the high numerical cost of the simulations for 3D population synthesis does not allow to follow the evolution of the discs up to very evolved stages where they can be considered isolated from the surrounding environment. MHD wind-driven disc populations can however be modelled in 1D using simple prescriptions as proposed by \cite{Suzuki+2016} or \cite{Tabone+2022MHDTheory}. Detecting characteristic signatures of either of the two evolutionary prescriptions is a compelling issue \citep{Long+2022, Alexander+2023, Somigliana+2023, Trapman+2023, Coleman+2024}.

In the context of the evolution of the correlations between the disc properties and the stellar mass, while the purely viscous scenario has been extensively studied by \cite{Somigliana+2022}, the wind-driven paradigm remains unexplored; with this paper, we address this deficiency and investigate the impact of MHD wind-driven evolution on the $M_{\mathrm{d}} - M_{\star}$ and $\dot M - M_{\star}$ correlations, with a particular focus on their time evolution and the comparison with the purely viscous paradigm. We also extend the work of \cite{Somigliana+2022} by including internal photoevaporation to the viscous framework. We employ numerical simulations of populations of protoplanetary discs, performed with the population synthesis code \texttt{Diskpop}, which we also introduce and release to the community. 

The paper is structured as follows: in Section \ref{sec:diskpop}, we present \texttt{Diskpop} and describe its main features, set up and solution algorithm; in Section \ref{sec:time_evo_analytical}, we discuss the time evolution of the $M_{\mathrm{d}} - M_{\star}$ and $\dot M - M_{\star}$ correlations in three evolutionary scenarios from the theoretical perspective; in Section \ref{sec:population_synth}, we show the impact of a spread in the initial conditions and dispersal mechanisms on the evolution of the slopes, and we present the numerical results obtained from realistic disc population synthesis; in Section \ref{sec:discussion} we interpret the implications of our findings in the context of the observational determination of the slopes, and finally in Section \ref{sec:conclusions} we draw the conclusions of this work.


\section{Numerical methods: \texttt{Diskpop}}\label{sec:diskpop}

In this Section we present the 1D population synthesis code \texttt{Diskpop}\footnote{\texttt{Diskpop} and the output analysis library \texttt{popcorn} can be installed via the Python Package Index, \texttt{pip install diskpop} and \texttt{pip install popcorn$\_$diskpop}. The full documentation and tutorials are available at https://alicesomigliana.github.io/diskpop-docs/index.html. If you use \texttt{Diskpop} in your work, please cite this paper (Somigliana et al. 2024).}. We describe the master equation for the secular evolution of discs (Section \ref{subsec:master_eq}), the initial conditions to generate a synthetic population (Section \ref{subsec:initcond}), the solution algorithm (Section \ref{subsec:algorithm}), and the user interface and output (Section \ref{subsec:output}). For a more detailed description, we refer to the code documentation; for a validation of the code, see Appendix \ref{sec:appendix_validation}.

\subsection{Master equation}\label{subsec:master_eq}

The master equation of protoplanetary disc evolution,

\begin{equation}
    \label{eq:master_equation}
        \begin{split}
            \frac{\partial \Sigma}{\partial t} = \frac{3}{r} \frac{\partial}{\partial r} \left[ \frac{1}{\Omega r} \frac{\partial}{\partial r} \left( r^2 \alpha_{\mathrm{SS}} \Sigma {c_s}^2 \right) \right] + \frac{3}{2r} \frac{\partial}{\partial r} \left[ \frac{\alpha_{\mathrm{DW}} \Sigma {c_s}^2}{\Omega} \right] \\
            - \frac{3 \alpha_{\mathrm{DW}} \Sigma {c_s}^2}{4 (\lambda-1)r^2 \Omega} - \dot{\Sigma}_{\mathrm{photo}},
        \end{split}
\end{equation}

\noindent describes the time evolution of the gas surface density in the most general framework, where $\Sigma$ is the gas surface density, $\Omega$ the Keplerian orbital frequency, $\alpha_{\mathrm{SS}}$ the \cite{ShakuraSunyaev1973} $\alpha$ parameter, $\alpha_{\mathrm{DW}}$ the MHD equivalent of $\alpha_{\mathrm{SS}}$ \citep{Tabone+2022MHDTheory}, $c_s$ the sound speed, and $\lambda$ the magnetic lever arm parameter, which quantifies the ratio of extracted to initial specific angular momentum. The four terms on the right hand side (RHS) refer to (i) the viscous torque, whose strength is parameterised by $\alpha_{\mathrm{SS}}$, (ii) the wind-driven accretion, which corresponds to an advection term, parameterised by $\alpha_{\mathrm{DW}}$, (iii) mass loss due to MHD disc winds, parameterised by $\lambda$ and (iv) mass loss due to other physical phenomena (in our case, we consider internal and external photoevaporation). Depending on the values of the specific parameters, Equation \eqref{eq:master_equation} can describe a purely viscous ($\alpha_{\mathrm{DW}} = 0$), purely MHD wind-driven ($\alpha_{\mathrm{SS}} = 0$) or hybrid ($\alpha_{\mathrm{SS}}, \alpha_{\mathrm{DW}} \neq 0$) evolution, with ($\dot \Sigma_{\mathrm{photo}} \neq 0$) or without ($\dot \Sigma_{\mathrm{photo}} = 0$) the influence of photoevaporation. In the following, we briefly describe the various evolutionary scenarios and the available analytical solutions.

\noindent \textbf{Viscously evolving discs.} In the case of purely viscous evolution, the MHD winds parameter $\alpha_{\mathrm{DW}}$ is set to zero. If we also neglect the influence of photoevaporation, Equation \eqref{eq:master_equation} reduces to the first term on the RHS and its solution depends on the functional form of the effective viscosity, parameterised as $\nu = \alpha_{\mathrm{SS}} c_s H$ (where $H$ is the vertical height of the disc). A popular analytical solution for viscous discs is the \cite{LyndenBellPringle1974} self-similar solution, which assumes viscosity to scale as a power-law of the radius ($\nu \propto R^{\gamma}$).

\noindent \textbf{MHD winds-driven evolution.} There are two classes of analytical solutions to Equation \eqref{eq:master_equation} in the MHD wind-driven scenario, associated with a specific prescription of $\alpha_{\mathrm{DW}}$ \citep{Tabone+2022MHDTheory}. We briefly describe their key features, and refer to the original paper for their derivation and an in-depth discussion.

\begin{enumerate}

    \item The simplest class of solutions (so-called \textit{hybrid solutions}), which highlight the main features of wind-driven accretion in comparison to the viscous model, assume a constant $\alpha_{\mathrm{DW}}$ with time; these solutions depend on the value of $\psi \equiv \alpha_{\mathrm{DW}}/\alpha_{\mathrm{SS}}$, which quantifies the relative strength of the radial and vertical torque. 

    \item Another class of solutions, which describe the unknown evolution of the magnetic field strength, assume a varying $\alpha_{\mathrm{DW}}$ with time. To obtain these, \cite{Tabone+2022MHDTheory} parameterised $\alpha_{\mathrm{DW}}(t) \propto \Sigma_{\mathrm{c}} (t)^{-\omega}$, with $\Sigma_{\mathrm{c}} = M_{\mathrm{d}}(t)/2 \pi {R_c}^2 (t)$ (where $R_c$ is a characteristic radius) and $\omega$ as a free parameter, and neglect the radial transport of angular momentum ($\alpha_{\mathrm{SS}} = 0$).
    
\end{enumerate}

\noindent \textbf{Photoevaporation.} The generic $\dot \Sigma_{\mathrm{photo}}$ term in Equation \eqref{eq:master_equation} allows to account for photoevaporative processes, both internal and external. The exact form of $\dot \Sigma_{\mathrm{photo}}$ depends on the specific model considered; therefore, the availability (or lack thereof) of analytical solutions needs to be considered case by case.

\texttt{Diskpop} allows to evolve populations of discs analytically. In particular, as of this release, it includes implementations of the \cite{LyndenBellPringle1974} self-similar solution and all the analytical solution proposed by \cite{Tabone+2022MHDTheory}. In the cases where Equation \eqref{eq:master_equation} cannot be solved analytically, the code relies on the solution algorithm described in Section \ref{subsec:algorithm}.

\subsection{Initial conditions and parameters}\label{subsec:initcond}

Every \texttt{Diskpop} simulation begins with the generation of a synthetic population of Young Stellar Objects (YSOs). Each YSO constitutes of a star and a disc, whose key initial parameters (stellar mass, disc mass, accretion rate, disc radius, evolutionary parameters $\alpha_{\mathrm{SS}}$, $\alpha_{\mathrm{DW}}$, $\lambda$, $\omega$...) can be set by the user. In the following, we describe the standard case where we consider the stellar masses to be distributed according to an Initial Mass Function (IMF) and correlating with the disc mass and radius, and briefly mention the other possible choices; for a deeper discussion, we refer to the \texttt{Diskpop} documentation.

\texttt{Diskpop} assembles YSOs by determining their parameters as follows:

\begin{enumerate}
    
    \item \textbf{Stellar mass $M_{\star}$:} determined following the \cite{Kroupa2001} IMF. Other possible choices are a constant mass for all the stars in the population, or a set of custom stellar masses.
    
    \item \textbf{Initial disc mass $M_{\mathrm{d}}$, accretion rate $\dot M$:} determined from log-normal distributions of given width and mean value. In the standard case, \texttt{Diskpop} considers an initial power-law correlation between the initial $M_{\mathrm{d}}$ and $\dot M$ and the stellar mass (see Section \ref{subsubsec:calculations_viscous} for a detailed discussion), where the normalisation at 1 M$_{\odot}$, the slope and the scatter around the power-laws are free parameters. If the correlations with the stellar mass are neglected, the user sets the mean value and spread of the distributions. 

    \item \textbf{Accretion parameters ($\alpha_{\mathrm{SS}}$, $\alpha_{\mathrm{DW}}$, $\lambda$, $\omega$):} global properties of the whole population, given as input from the user. By setting the parameters controlling accretion, \texttt{Diskpop} determines the disc radius $R_{\mathrm{d}}$ and accretion timescale $t_{\mathrm{acc}}$ - which are instead disc-specific and linked to the disc mass and accretion rate.

    \item \textbf{Internal photoevaporation parameters ($\dot M_{\mathrm{wind}}$, $L_X$):} the total photoevaporative mass-loss rate, $\dot M_{\mathrm{wind}}$, can either be set by the user or computed from the stellar X luminosity $L_X$ as \citep{Owen+2012}
    
    \begin{equation*}
        \dot M_{\mathrm{wind}} = 6.25 \times 10^{-9} \times \left( \frac{M_{\star}}{M_{\odot}} \right)^{-0.068} \left( \frac{L_X}{10^{30} \mathrm{erg}\mathrm{s}^{-1}} \right)^{1.14} M_{\odot} \mathrm{yr}^{-1}.
    \end{equation*}

    \noindent The surface mass-loss profile $\dot \Sigma_{\mathrm{photo}}$ (Equation B2 in \citealt{Owen+2012}) is then scaled so that $\int 2 \pi R \dot \Sigma_{\mathrm{photo}} \rm{d}R$ is equal to $\dot M_{\mathrm{wind}}$. Like for $M_{\mathrm{d}}$ and $\dot M$, $\dot M_{\rm{wind}}$ (or equivalently $L_X$) is extracted from a log-normal distribution whose mean is determined assuming power-law correlations with the stellar mass, while the normalisation at 1 M$_{\odot}$, the slope and the width of the distribution are free parameters.
    
    \item \textbf{External photoevaporation parameters ($FUV$):} FUV flux experienced by each disc, in units of G$_{0}$\footnote{G$_{0}$ stands for the Habing unit \citep{Habing_g0}, the flux integral over the range of wavelengths [912 - 2400] $\rm{\mathring{A}}$ weighted by the average value in the solar neighbourhood (1.6$\times 10^{-3}$ erg s$^{-1}$ cm$^{-2}$).}. This parameter can be set to any value accessible in the FRIEDv2 grid of mass loss rate \citep{Friedv2}, spanning from 1 to $10^{5}$ G$_{0}$.
    
\end{enumerate}

\subsection{Solution algorithm}\label{subsec:algorithm}

After generating the initial population of YSOs as described above, \texttt{Diskpop} proceeds to evolve it by integrating the master equation \eqref{eq:master_equation}. Our solution algorithm employs an operator splitting method: the original equation is separated into different parts over a time step, and the solution to each part is computed separately. Then, all the solutions are combined together to form a solution to the original equation. We split Equation \eqref{eq:master_equation} into five different pieces, related to viscosity, wind-driven accretion onto the central star, wind-driven mass loss, internal and external photoevaporation respectively. Furthermore, \texttt{Diskpop} includes the possibility to trace the dust evolution in the disc, which is split in radial drift and dust diffusion. In the following, we describe the solution algorithm for each process.

\begin{enumerate}
 \setlength{\itemsep}{2pt}

    \item \textbf{Viscous accretion:} the standard viscous solver is based on the freely available code by \cite{Booth+2017-viscouscode}. We assume a radial temperature profile $T \propto R^{-1/2}$, which results in $c_{s} \propto R^{-1/4}$ and $H/R \propto R^{1/4}$. Note that this implies $\nu \propto R$ (i.e., $\gamma = 1$), which will be the case from now on. We assume $H/R = 1/30$ at 1 AU and a mean molecular weight of $2.4$. We refer to the original paper for details on the algorithm.
    
    \item \textbf{Wind-driven accretion:} the second term in Equation \eqref{eq:master_equation} is effectively an advection term. The general form of the advection equation for a quantity $q$ with velocity $v$ is $\partial_t q(x, t) + v \partial_x q(x, t) = 0$; in the case of wind-driven accretion, the advected quantity is $R \Sigma$, while the advection (inwards) velocity is given by $v_{\mathrm{DW}} = (3 \alpha_{\mathrm{DW}} H c_s)/2R$. We solve the advection equation with an explicit upwind algorithm (used also for dust radial drift).

    \item \textbf{Wind-driven mass loss:} the mass loss term (third in Equation \ref{eq:master_equation}) does not involve any partial derivative, and therefore is simply integrated in time multiplying by the time step.

    \item \textbf{Internal photoevaporation:} effectively, internal photoevaporation (implemented through the model of \citealt{Owen+2012}) is another mass loss term - therefore, as above, its contribution is computed with a simple multiplication by the time step. Once the accretion rate of the disc drops below the photoevaporative mass loss rate, a gap opens in the disc at the radius of influence of photoevaporation: in the model of \cite{Owen+2012}, the prescription changes depending on the radial location in the disc, with respect to the gap itself. Later, the gap continues to widen; when it eventually becomes larger than the disc, we stop the evolution and consider the disc as dispersed.

    \item \textbf{External photoevaporation:} for a given stellar mass and FUV flux experienced by the disc, the mass loss rate arising from external photoevaporation is obtained, at each radial position, from a bi-linear interpolation of the FRIEDv2 grid \citep{Friedv2} using the disc surface density at each radial cell. The outside-in depletion of material is implemented following the numerical approach of \citet{Sellek+2020-dustyorigin}: we define the \textit{truncation} radius, $R_{\mathrm{t}}$, as the position in the disc corresponding to the maximum photoevaporation rate (which is related to the optically thin/thick transition of the wind), and we remove material from each grid cell at $R>R_{\mathrm{t}}$ weighting on the total mass outside this radius. The mass loss attributed to the cell $i$ can be written as:
    \begin{equation}
        \dot{M}_{\mathrm{ext},i} = \dot{M}_{\mathrm{tot}} \frac{M_{i}}{M(R>R_{\mathrm{t}})},
    \end{equation}
    where $M_{i}$ is the mass contained in the cell $i$, and $\dot{M}_{\mathrm{tot}}$ is the total mass loss rate outside the truncation radius.

    \item \textbf{Dust evolution\footnote{As the dust evolution module was forked from Richard Booth's repository, users of \texttt{Diskpop} who wish to use dust in their work ought to cite \cite{Booth+2017-viscouscode} together with this paper.}}: based on the two populations model by \cite{Birnstiel2012} and the implementation of \cite{Booth+2017-viscouscode}. We consider the dust grain distribution to be described by two representative sizes, a constant monomer size and a time-dependent larger size, which can grow up to the limit imposed by the fragmentation and radial drift barriers. We evolve the dust fraction of both sizes following \cite{Laibe&Price2014-onefluid}, and also include a diffusive term: the diffusion comes from the coupling with the turbulent gas, which has the effect of mixing the dust grains, counteracting gradients in concentration \citep{Birnstiel2010}. The dust-gas relative velocities are computed following \cite{Tanaka+2005} and include feedback on the gas component. We refer to \cite{Booth+2017-viscouscode} for details on the numerical implementation. Dust evolution is included in the release of \texttt{Diskpop}, however the scientific results presented in this work are based on gas simulations only.
    
\end{enumerate}

The separate pieces of Equation \eqref{eq:master_equation} must be solved over the same time step to be joined in a coherent solution. We calculate the time step for each process imposing the Courant-Friedrichs-Lewy (CFL) condition. The CFL condition reads $\Delta t = C \mathrm{ min}(\Delta x / v)$ and ensures that, within one time step $\Delta t$, the material moving at velocity $v$ does not flow further than one grid spacing $\Delta x$. The Courant number $C$ must be positive and smaller than 1, with $C = 1$ corresponding to the maximum allowed timestep to keep the algorithm stable. In our implementation, we pick $C = 0.5$. We use zero gradients boundary conditions, setting the value of the first and last cell in our grid to that of the second and second to last. We solve the equation on a radial grid of $10^3$ points with power-law spacing and exponent $1/2$, extending from $3 \times 10^{-3}$ au to $10^4$ au. From the physical point of view, this choice corresponds to assuming boundary layer accretion (see., e.g., \citealt{Popham+1993, Kley&Lin1996}) - however the difference from magnetic truncation accretion is negligible beyond $\sim 10^{-3}$ au.

After each process has been solved separately, all the pieces are put back together to compute the new surface density, from which the integrated disc quantities are then calculated. As each disc evolves independently of the others in the population, the solver can easily be run in parallel.

\subsection{User interface and output}\label{subsec:output}

The user interface of \texttt{Diskpop} is a .json parameters file which includes all the user-dependent parameters. Aside from the number of objects in the population and the evolutionary mechanism, the user can set the chosen IMF (either \citealt{Kroupa2001}, single stellar mass, or custom input file), the distributions to draw the disc parameters from (single value, flat, normal, log-normal), as well as the normalisation, slope and spread of the correlations, the times at which snapshots are generated, and the initial dust-to-gas ratio. Furthermore, the user can determine a limit disc mass: this is to be intended as a threshold below which the disc would not be detectable anymore, and is therefore considered dispersed in the simulation as well. When a disc is dispersed, the corresponding YSO turns into a Class III object consisting of the central star only.

The output of \texttt{Diskpop} is a .hdf5 file containing the properties of both the disc and the star at all chosen time steps for each YSO in the population: this includes the stellar mass, luminosity, temperature, disc mass, accretion rate, accretion timescale, gas and dust surface density, disc radius, dust grain sizes. The output can be easily read and analysed with the dedicated library \texttt{popcorn}, released with the code. For a more in-depth description of the parameters, the user interface and the output, we refer to the \texttt{Diskpop} documentation.


\section{The time evolution of the correlations between disc properties and stellar mass under different accretion drivers: analytical considerations}\label{sec:time_evo_analytical}

\begin{table}[h]
    \centering
    \vspace{0.2cm}

    \begin{tabular}[H]{P{1.5cm} P{5.5cm}}
    
    \hline

    \rule{0pt}{2.5ex} Parameter & Description \\[1ex]

    \hline
    
    \rule{0pt}{2.5ex} $\lambda$ & \footnotesize{Magnetic lever arm parameter} \\
    $\psi$ & \footnotesize{Wind-to-turbulent $\alpha$ ratio} \\
    $\omega$ & \footnotesize{Power-law index of $\alpha_{\mathrm{DW}}$ with $\Sigma_{\mathrm{c}}$} \\
    $\xi$ & \footnotesize{Mass ejection index} \\
    $f_{\mathrm{M}, 0}$ & \footnotesize{Initial mass ejection-to-accretion ratio} \\[1ex]
    
    \hline
    
    \rule{0pt}{2.5ex} $\lambda_{\mathrm{m}}$ & \footnotesize{Power-law index of $M_{\mathrm{d}}$ with $M_{\star}$} (\footnotesize{Eq. \ref{eq:assumptions}}) \\
    $\lambda_{\mathrm{acc}}$ & \footnotesize{Power-law index of $\dot M$ with $M_{\star}$} (\footnotesize{Eq. \ref{eq:assumptions}}) \\
    $\zeta$ & \footnotesize{Power-law index of $R_{\mathrm{d}}$ with $M_{\star}$} (\footnotesize{Eq. \ref{eq:assumptions}}) \\
    $\beta$ & \footnotesize{Power-law index of $H/R$ with $M_{\star}$} \\
    $\mu$ & \footnotesize{Power-law index of $t_{\mathrm{acc}}$ with $M_{\star}$} \\
    $\delta$ & \footnotesize{$\lambda_{\mathrm{m}} - \lambda_{\mathrm{acc}}$} \\[0.5ex]

    \hline

    \end{tabular}
    
    \hspace{1.5cm}
    
    \caption{Summary and description of the parameters used throughout the paper: the top block refers to the MHD parameters defined in \cite{Tabone+2022MHDTheory}, while the bottom block shows the slopes of the correlations between the disc properties and the stellar mass.}
    \label{tab:params_summary}
\end{table}

The existence of power-law correlations between the main integrated disc properties - namely the disc mass and stellar accretion rate - and the stellar mass is supported by various surveys across a number of different star-forming regions (e.g., on $M_{\mathrm{d}}-M_{\star}$: \citealt{Ansdell+2016Survey, Barenfeld2016-masssurvey, Pascucci+2016, Testi+2016}; on $\dot M - M_{\star}$: \citealt{Muzerolle+2003, Natta+2004-lowmassaccrates, Mohanty+2005, Alcala+2014, Manara+2016-ChaI, Alcala+2017, Manara2017-Cha, Venuti+2019accrates, Manara+2020-UpperSco, Testi+2022-Ofiucone}). However, whether the establishment and subsequent evolution of said correlations is a product of the secular evolution of discs, or rather an imprint of the initial conditions, remains unclear. \cite{Somigliana+2022} explored a combination of both possibilities, assuming the correlations to hold as initial conditions and investigating the impact of purely viscous evolution; we briefly recall their main theoretical results (Section \ref{subsubsec:calculations_viscous}) and extend their analysis to the hybrid (Section \ref{subsubsec:calculations_MHD_omega0}) and purely wind-driven (Section \ref{subsubsec:calculations_MHD_omeganot0}) models from the theoretical perspective. We note that the results presented in this work are based on gas simulations.

Following \cite{Somigliana+2022}, we assume power-law correlations between the disc properties and the stellar mass to hold as \textit{initial conditions}. We focus on the disc mass $M_{\mathrm{d}}$, the stellar accretion rate $\dot M$ and the disc radius $R_{\mathrm{d}}$, and label the slopes of their correlations with the stellar mass $\lambda_{\mathrm{m}}$, $\lambda_{\mathrm{acc}}$, and $\zeta$ respectively. The initial correlations are set as follows:
    
    \begin{equation}
        \begin{cases}
            M_\mathrm{d} (0) \propto {M_{\star}}^{\lambda_{\mathrm{m}, 0}}, \\
            \dot{M} (0) \propto {M_{\star}}^{\lambda_{\mathrm{acc}, 0}}, \\
            R_{\mathrm{d}} (0) \propto {M_{\star}}^{\zeta_0}.
        \end{cases}
    \label{eq:assumptions}
    \end{equation}

To analyse the impact of secular evolution on this set of initial conditions, we analytically determine the evolved expressions for $M_\mathrm{d} (t)$, $\dot{M} (t)$ and $R_{\mathrm{d}} (t)$ in the three different scenarios. Table \ref{tab:params_summary} summarises the parameters introduced in this Section.


\subsection{Purely viscous model}\label{subsubsec:calculations_viscous}

The full calculations for the purely viscous case can be found in \cite{Somigliana+2022}. Here, we briefly remind the main assumptions and results, and we refer to the original paper for a detailed discussion.

As mentioned in Section \ref{subsec:master_eq}, assuming a power-law scaling of viscosity with the disc radius ($\nu \propto R^{\gamma}$) allows to solve the viscous evolution equation analytically, recovering the so-called self-similar solution \citep{LyndenBellPringle1974}. In this case, the disc mass and accretion rate read

\begin{equation}
    M_\mathrm{d} (t) = M_{\mathrm{d}, 0} \left( 1 + \frac{t}{t_{\nu}} \right)^{1 - \eta},
    \label{eq:discmass_ss}
\end{equation}

\begin{equation}
    \dot M (t) = (\eta - 1) \frac{M_{\mathrm{d}, 0}}{t_{\nu}} \left( 1 + \frac{t}{t_{\nu}} \right)^{- \eta},
    \label{eq:accrate_ss}
\end{equation}

\noindent where $\eta = (5/2 - \gamma)/(2 - \gamma)$ and the viscous timescale $t_{\nu} = {R_c}^2/[3(2-\gamma)^2 \nu(R = R_c)]$ at the characteristic radius $R_c$. Because $\dot M_0 \propto M_{\mathrm{d}, 0}/t_{\nu, 0}$, a power-law scaling of $M_{\mathrm{d}, 0}$ and $\dot M_0$ with the stellar mass implies the viscous timescale $t_{\nu, 0}$ to scale as a power-law with the stellar mass as well, which we define as $t_{\nu, 0} \propto {M_{\star}}^{\mu_0}$; furthermore, this scaling corresponds to the difference between the scaling of the disc mass with the stellar mass and of the accretion rate with the stellar mass. Defining $\delta_0 = \lambda_{\mathrm{m}, 0} - \lambda_{\mathrm{acc}, 0}$, in this case $\mu_0 = \delta_0$\footnote{The definition of $\delta_0$ might seem redundant at this stage, but it will become important in the following discussion.}; therefore, the scaling of $t_{\nu, 0}$ with the stellar mass is determined by the relative values of $\lambda_{\mathrm{m}, 0}$ and $\lambda_{\mathrm{acc}, 0}$. The main results of \cite{Somigliana+2022} are that (i) viscous evolution maintains the power-law shape of the correlations between the stellar mass and the disc parameters, however (ii) the slope of said correlations may evolve with time, depending on the initial conditions. This is because in a purely viscous framework, the $M_{\mathrm{d}}-\dot M$ correlation is bound to reach a linear correlation with slope unity \citep{Lodato2017, Rosotti+2017}, which implies the two quantities to have the same dependence on the stellar mass, as $M_{\mathrm{d}} / \dot M \propto {M_{\star}}^{\lambda_{\mathrm{m}} - \lambda_{\mathrm{acc}}}$. Therefore, $\lambda_{\mathrm{m}}$ and $\lambda_{\mathrm{acc}}$ must eventually reach the same value, determined by the initial conditions as

\begin{equation}
    \lambda_{\mathrm{m, evo}} = \lambda_{\mathrm{acc, evo}} = \frac{3 \lambda_{\mathrm{m}, 0} - \lambda_{\mathrm{acc}, 0}}{2}.
    \label{eq:lambdam=lambdaac_evolved}
\end{equation}

\noindent Depending on the sign of $\delta_0 = \lambda_{\mathrm{m}, 0} - \lambda_{\mathrm{acc}, 0}$, the initial slopes can either

\begin{itemize}
    \item \textit{steepen}, i.e. $\lambda_{\mathrm{m, evo}} > \lambda_{\mathrm{m}, 0}$, if $\delta_0 > 0$ (implying also
    
    \noindent $\lambda_{\mathrm{acc, evo}} > \lambda_{\mathrm{acc}, 0}$);
    \item \textit{flatten}, i.e. $\lambda_{\mathrm{m, evo}} < \lambda_{\mathrm{m}, 0}$, if $\delta_0 < 0$ (implying also 
    
    \noindent $\lambda_{\mathrm{acc, evo}} < \lambda_{\mathrm{acc}, 0}$);
    \item \textit{remain constant}, i.e. $\lambda_{\mathrm{m, evo}} = \lambda_{\mathrm{m}, 0}$, if $\delta_0 = 0$ (implying 
    
    \noindent also $\lambda_{\mathrm{acc, evo}} = \lambda_{\mathrm{acc}, 0}$).
\end{itemize}

\noindent Because in the viscous case $\delta_0 = \mu_0$, where we note again that $\mu_0$ is the slope of the correlation between the viscous timescale and the stellar mass ($t_{\nu, 0} \propto {M_{\star}}^{\mu_0}$), we can also interpret these scenarios from the viscous timescale perspective. If $\mu_0 > 0$, meaning that the viscous timescale increases with the stellar mass, discs around less massive stars will have shorter viscous timescales, which leads to a faster evolution, compared to discs around more massive stars, which will in turn have longer viscous timescales. This uneven evolution across the stellar mass spectrum leads to a steepening of the linear correlation, as is visualised by \cite{Somigliana+2022} in Figure 1. The same reasoning, but with opposite/constant trend, applies to the other two scenarios.



\subsection{Hybrid model - \texorpdfstring{$\omega = 0$}{omega = 0}}\label{subsubsec:calculations_MHD_omega0}

In the hybrid viscous and MHD winds model, the general analytical solution by \cite{Tabone+2022MHDTheory} gives 

\begin{equation}
    M_\mathrm{d} (t) = M_0 \left( 1 + \frac{t}{(1 + \psi) t_{\mathrm{acc}, 0}} \right)^{-(\psi + 2 \xi + 1)/2},
\label{eq:mdisc_mhd_omegazero}
\end{equation}

\begin{equation}
    \dot{M} (t) = \dot{M}_0 \left( 1 + \frac{t}{(1 + \psi) t_{\mathrm{acc}, 0}} \right)^{-(\psi + 4 \xi + 3)/2},
\label{eq:mdot_mhd_omegazero}
\end{equation}

\noindent where $\dot{M}_0$ is defined as

\begin{equation}
    \dot{M}_0 = \frac{\psi + 1 + 2 \xi}{\psi + 1} \frac{M_0}{2 t_{\mathrm{acc}, 0}} \frac{1}{(1+f_{\mathrm{M}_0})};
\label{eq:mdot0_mhd_omegazero}
\end{equation}

\begin{figure*}
    \centering
    \includegraphics[width = \textwidth]{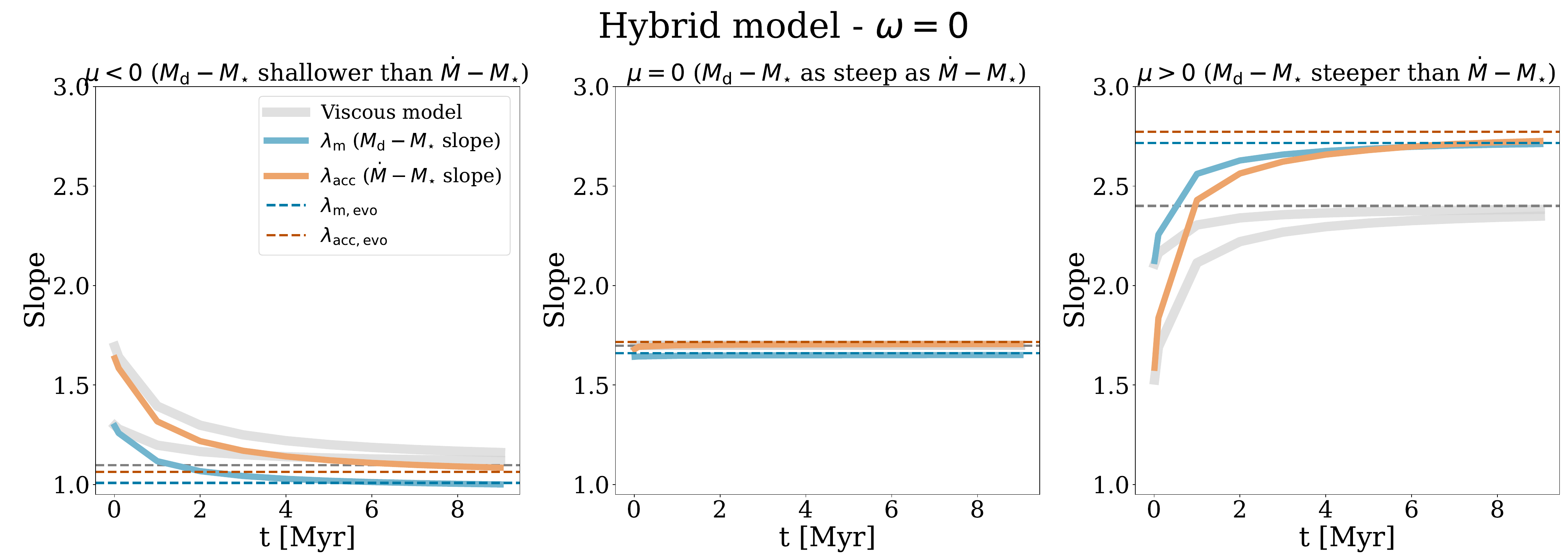}
    \caption{Time evolution of the slopes of the $M_{\mathrm{d}}-M_{\star}$ and $\dot M-M_{\star}$ correlations, $\lambda_{\mathrm{m}}$ and $\lambda_{\mathrm{acc}}$ (blue and orange solid line respectively) in the hybrid scenario with $\alpha_{\mathrm{SS}} = \alpha_{\mathrm{DW}} = 10^{-3}$ ($\psi = 1$), $\lambda = 3$, $\beta = 0.5$, resulting in $\xi = 0.11$. The dashed lines represent the expected evolved value of both slopes, as in Equation \eqref{eq:lambda_evo_mhd}. For comparison, we include the viscous evolution as well, represented by grey solid (actual evolution) and dashed (expected evolved value) lines (see \citealt{Somigliana+2022} for a detailed discussion). The three panels show different values of $\mu_0$, slope of the $t_{\mathrm{acc}, 0} - M_{\star}$ correlation, which is directly linked to the difference between $\lambda_{\mathrm{m}, 0}$ and $\lambda_{\mathrm{acc}, 0}$ (see text for details): we expect the slopes to (i) decrease if $\mu_0 < 0$ (left panel), (ii) remain constant if $\mu_0 = 0$ (central panel), and (iii) increase if $\mu_0 > 0$ (right panel). Contrary to the viscous case, the slopes are not expected to reach the same value anymore, but rather settle to a constant difference of $- \xi/2$. This difference is always negative, meaning that the evolved $\dot M - M_{\star}$ correlation is always steeper than that of $M_{\mathrm{d}} - M_{\star}$ (explaining the lines crossing in the right panel).}
    \label{fig:slopes_mhd_omegazero}
\end{figure*}

\noindent in this notation, $\psi = \alpha_{\mathrm{DW}}/\alpha_{\mathrm{SS}}$ represents the relative strength of MHD winds and viscosity, 

\begin{equation*}
    \xi = \frac{1}{4} (\psi + 1) \left[ \sqrt{1 + \frac{4\psi}{(\lambda-1)(\psi + 1)^2}} - 1 \right]
\end{equation*}

\noindent is the mass ejection index quantifying the local mass loss rate to the local accretion rate, and $f_{\mathrm{M}, 0} = (R_{c, 0}/R_{in})^{\xi} - 1$ the dimensionless mass ejection-to-accretion ratio (with $R_{in}$ initial disc radius). If we neglect the MHD-driven mass loss ($\psi \ll 1$ and $\xi \ll 1$, which correspond to $f_{\mathrm{M}, 0} \ll 1$ as well), Equations \eqref{eq:mdisc_mhd_omegazero} and \eqref{eq:mdot_mhd_omegazero} reduce to the viscous case; on the other hand, if mass loss is included, it depends on the radial extent of the disc through $f_{\mathrm{M}, 0} + 1$ - which has an impact on the initial accretion rate (Equation \ref{eq:mdot0_mhd_omegazero}). Because the accretion timescale $t_{\mathrm{acc}}$ is a generalisation of $t_{\nu}$ in the MHD winds framework, the dependence of the two timescales on the stellar mass is exactly equivalent, and we will keep the same notation as above: $t_{\mathrm{acc}} \propto {M_{\star}}^{\mu}$. However, as mentioned above $\dot M_0$ depends on the stellar mass not only through $M_0$ and $t_{\mathrm{acc}, 0}$ as in the viscous case, but also through $f_{\mathrm{M},0} + 1$. As $f_{\mathrm{M},0} + 1 \propto {R_{c, 0}}^{\xi}$, and $R_{c, 0} \propto {M_{\star}}^{\zeta_0}$, the additional dependence will have a slope of $\zeta_0 \xi$. Therefore, in the MHD winds scenario we can link $\delta_0$ with $\mu_0$ as $\delta_0 = \mu_0 + \zeta_0 \xi$. The practical meaning of this difference is that, while in the viscous scenario the difference in slope between the two correlations depends only on the scaling of the viscous timescale with the stellar mass, in the hybrid scenario it depends also on the scaling between the disc radius and the stellar mass. It is important to note that $\xi$ is a small number, typically of the order of $\sim 0.1$, therefore the difference between the viscous and hybrid case is not particularly prominent. For evolved populations, the disc mass and accretion rate read \citep{Tabone+2022MHDTheory}

\begin{equation}
    M_\mathrm{d} (t \gg t_{\mathrm{acc}}) \sim M_0 \left(\frac{t}{t_{\mathrm{acc}, 0}} \right)^{-(\psi + 2 \xi + 1)/2},
\label{eq:mdisc_mhd_omegazero_evo}
\end{equation}

\begin{equation}
    \dot{M} (t \gg t_{\mathrm{acc}}) \sim \dot{M}_0 \left(\frac{t}{t_{\mathrm{acc}, 0}} \right)^{-(\psi + 4 \xi + 3)/2};
\label{eq:mdot_mhd_omegazero_evo}
\end{equation}

\noindent this brings the evolved slopes $\lambda_{\mathrm{m, evo}}$ and $\lambda_{\mathrm{acc, evo}}$ to

\begin{equation}
    \begin{split}
        \lambda_{\mathrm{m, evo}} = \lambda_{\mathrm{m}, 0} + \frac{1}{2} \mu_0 (\psi + 2\xi + 1), \\
        \lambda_{\mathrm{acc, evo}} = \lambda_{\mathrm{acc}, 0} + \frac{1}{2} \mu_0 (\psi + 4\xi + 3),
    \end{split}   
    \label{eq:lambda_evo_mhd}
\end{equation}

\begin{figure*}
    \centering
    \includegraphics[width = \textwidth]{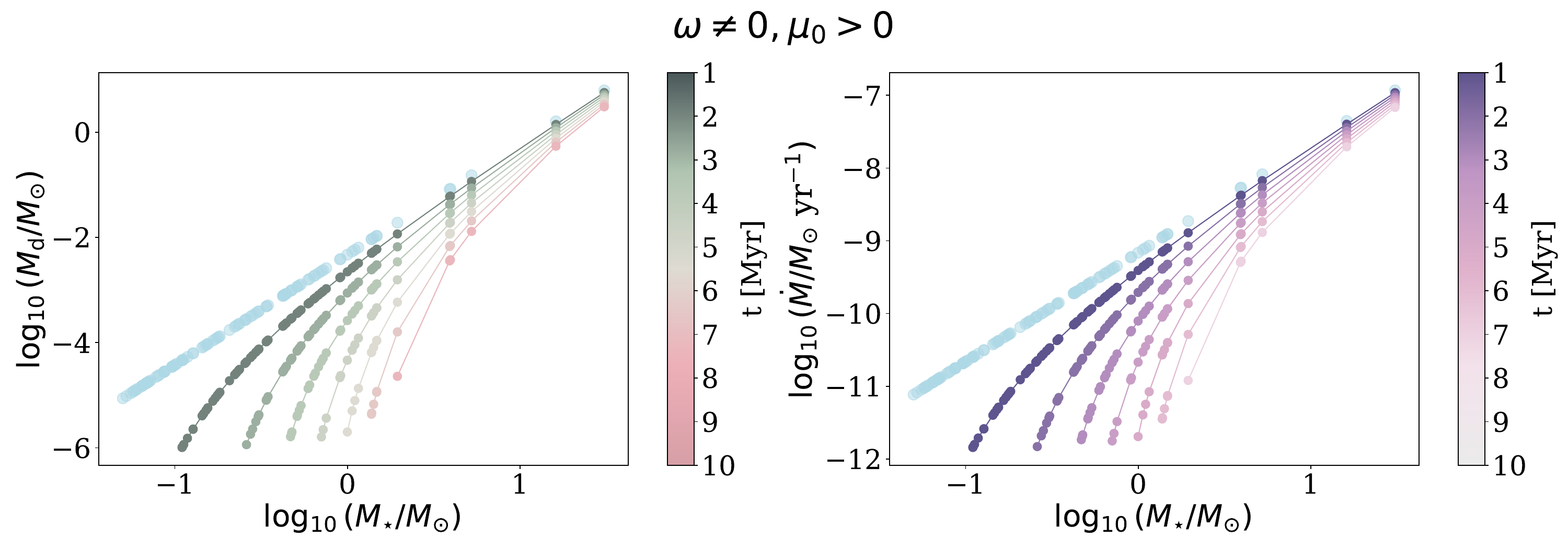}
    \caption{Time evolution of $M_{\mathrm{d}} - M_{\star}$ (left) and $\dot M - M_{\star}$ (right) in the pure wind model ($\omega \neq 0$). This plot is obtained with disc population synthesis modelling, without any spread in the initial conditions. Each dot represents a disc in the population at different ages as shown in the colour bar. The initial power-law correlation, shown in light blue, is lost as early as $\sim 1$ Myr (corresponding to $\sim 2 <t_{\mathrm{acc}, 0}>$ with these parameters) due to a downward bending corresponding to lower stellar masses. In this simulation, we have used $N = 100$ discs, $\alpha_{\mathrm{DW}} = 10^{-3}$, $\lambda = 3$, $\omega = 0.25$, $\lambda_{\mathrm{m}, 0} = 2.1$, $\lambda_{\mathrm{acc}, 0} = 1.5$. The set of MHD parameters is based on \citealt{Tabone+2022MHDTheory}.}
    \label{fig:brokencorr_MHD}
\end{figure*}

\noindent which reduces to Equation \eqref{eq:lambdam=lambdaac_evolved} in the viscous case ($\psi \ll 1$, $\xi \ll 1$). Like viscosity, a hybrid secular evolution maintains the power-law shape of the correlation; moreover, Equation \eqref{eq:lambda_evo_mhd} provides a theoretical prediction for the evolved slopes: they can steepen, flatten or remain the same as the initial conditions, depending on the involved parameters. As the terms in parentheses in Equation \eqref{eq:lambda_evo_mhd} are sums of positive values, the sign of the evolved slopes depends on the sign of $\mu_0$ like in the viscous case. However, there is a difference from the viscous case: as in the hybrid scenario $\mu_0$ and $\delta_0$ do not coincide anymore, a constraint on the value of $\mu_0$ is translated into a constraint on $\delta_0 - \zeta_0 \xi$. In particular, the slopes will increase if $\delta_0 > \zeta_0 \xi$ (corresponding to $\mu_0 > 0$), whereas if $\delta_0 < \zeta_0 \xi$ (corresponding to $\mu_0 < 0$), the slopes will decrease; and finally, if $\delta_0 = \zeta_0 \xi$ (corresponding to $\mu_0 = 0$) the slopes will remain constant in time. Another difference from the viscous scenario is that $\lambda_{\mathrm{m, evo}}$ and $\lambda_{\mathrm{acc, evo}}$ are not expected to reach the same value anymore. The limit difference is given by $\delta_{\mathrm{evo}} = \lambda_{\mathrm{m, evo}} - \lambda_{\mathrm{acc, evo}}$: substituting the values from Equation \eqref{eq:lambda_evo_mhd} one finds $\delta_{\mathrm{evo}} = \delta_0 - \mu_0 (\xi + 1)$, which can be further reduced to $\delta_{\mathrm{evo}} = \xi (2 \beta + \frac{1}{2})$ by using the definition of $\mu_0$ as the slope of the $t_{\mathrm{acc}, 0} - M_{\star}$ correlation. In this expression, $\beta$ is the slope of the correlation between the disc aspect ratio and the stellar mass, which comes from definition of the accretion timescale; with a standard $\beta = -1/2$ (a reasonable approximation of the value derived by radiative transfer simulation, see e.g., \citealt{Sinclair2020}), we obtain $\delta_{\mathrm{evo}} = -\xi/2$. As $\xi$ is positive by definition, in the hybrid scenario $\delta_{\mathrm{evo}}$ is always negative, meaning that in an evolved population, the correlation between the accretion rate and the stellar mass will necessarily be steeper than that between the disc mass and the stellar mass. However, we stress once more that $\xi$ is a small number and therefore the predicted difference is also small. Figure \ref{fig:slopes_mhd_omegazero} shows the evolution of the slopes from \texttt{Diskpop} simulations (with no spread in the initial conditions) for the hybrid model (coloured) compared with the viscous case (grey), which matches the theoretical expectations discussed above. The list of parameters used in the simulations is available in Table \ref{tab:params}.

Summarising, both the hybrid and viscous secular evolution preserve the power-law shape of the correlations between the disc properties and the stellar mass. The main difference is that the hybrid model does not predict the slopes of the $M_{\mathrm{d}} - M_{\star}$ and $\dot M - M_{\star}$ correlations to reach the same limit value (unlike the viscous case). The predicted difference in the evolved slopes is given by $\delta_{\mathrm{evo}} = -\xi/2$ (of the order of 0.1). The current observational uncertainties on $\delta_{\mathrm{evo}}$ range from 0.4 to 0.8 (\citealt{Testi+2022-Ofiucone}, see Table \ref{tab:delta_Testi}), 4 to 8 times larger than the predicted difference, making it not observable at this stage.

\subsection{Pure wind - $\omega \neq 0$}\label{subsubsec:calculations_MHD_omeganot0}

The solution to the pure wind model (i.e., time-dependent $\alpha_{\mathrm{DW}}$ through $\alpha_{\mathrm{DW}}(t) \propto \Sigma_{\mathrm{c}}(t)^{- \omega}$) by \cite{Tabone+2022MHDTheory} gives

\begin{equation}
    M_\mathrm{d} (t) = M_0 \left( 1 - \frac{\omega}{2 t_{\mathrm{acc}, 0}} t \right)^{1/\omega},
\label{eq:mdisc_mhd_omeganotzero}
\end{equation}

\begin{equation}
    \dot{M} (t) = \frac{M_0}{2 t_{\mathrm{acc}, 0} (1+f_{\mathrm{M}, 0})} \left( 1 - \frac{\omega}{2 t_{\mathrm{acc}, 0}} t \right)^{-1+1/\omega};
\label{eq:mdot_mhd_omeganotzero}
\end{equation}

\noindent in this case, the functional form of Equation \eqref{eq:mdisc_mhd_omeganotzero} and \eqref{eq:mdot_mhd_omeganotzero} does not allow us to derive a simple analytical expression for $M_\mathrm{d} (t \gg t_{\mathrm{acc}})$ and $\dot M (t \gg t_{\mathrm{acc}})$. Therefore, to explore the evolution of the correlations in the pure wind case, we fully rely on \texttt{Diskpop} simulations. In order to account for the impact of secular evolution only, we input perfect correlations between the disc properties and the stellar mass - that is, we do not include any spread in the initial conditions. Figure \ref{fig:brokencorr_MHD} shows the time evolution of $M_{\mathrm{d}}$ (left panel) and $\dot M$ (right panel) as a function of the stellar mass, from younger (darker) to older (lighter) populations. The input power-law correlation (light blue) corresponds to a line in the logarithmic plane; however, as early as $\sim 1$ Myr (corresponding to $\sim 2<t_{\mathrm{acc}, 0}>$ for this simulation), the input correlation starts to bend downwards at lower stellar masses. This behaviour reveals a significantly different trend from the viscous and hybrid model: in the pure wind scenario, the initial power-law shape of the correlations is not preserved by the secular evolution, but rather broken. In Figure \ref{fig:brokencorr_MHD}, the faster evolution of discs around lower mass stars is the consequence of a positive $\mu_0$, implying a positive correlation between the stellar mass and the accretion timescale; a negative correlation between $M_{\star}$ and $t_{\mathrm{acc}, 0}$ (i.e., a negative $\mu_0$) would lead to faster evolution of discs around higher mass stars, causing the correlation to bend towards the other end of the stellar mass spectrum (see Figure \ref{fig:appendix_brokencorr_MHD}). Another parameter that might impact the evolution of the correlation is the dispersal timescale, $t_{\mathrm{disp}}$; in the wind-driven case, $t_{\mathrm{disp}} \propto t_{\mathrm{acc}}$, therefore $t_{\mathrm{disp}}$ does not introduce any further dependence on the stellar mass.

\section{Population synthesis}\label{sec:population_synth}

In Section \ref{sec:time_evo_analytical} we have discussed analytical trends, and presented simulations with no spread to analyse the effect of secular disc evolution alone on the evolution of the slopes. In order to test our theoretical predictions against observational data, we need to account for both a spread in the initial conditions and disc dispersal mechanisms. In this Section, we discuss the impact of both factors on the $M_{\mathrm{d}} - M_{\star}$ and $\dot M - M_{\star}$ slopes and run realistic population synthesis simulations, to determine whether the model-dependent evolutionary features described in Section \ref{sec:time_evo_analytical} would be observable with the currently available data.

\subsection{Effects of a spread in the initial conditions}\label{subsec:effectsofinitcond}

\begin{figure*}[h]
    \centering
    \includegraphics[width = \textwidth]{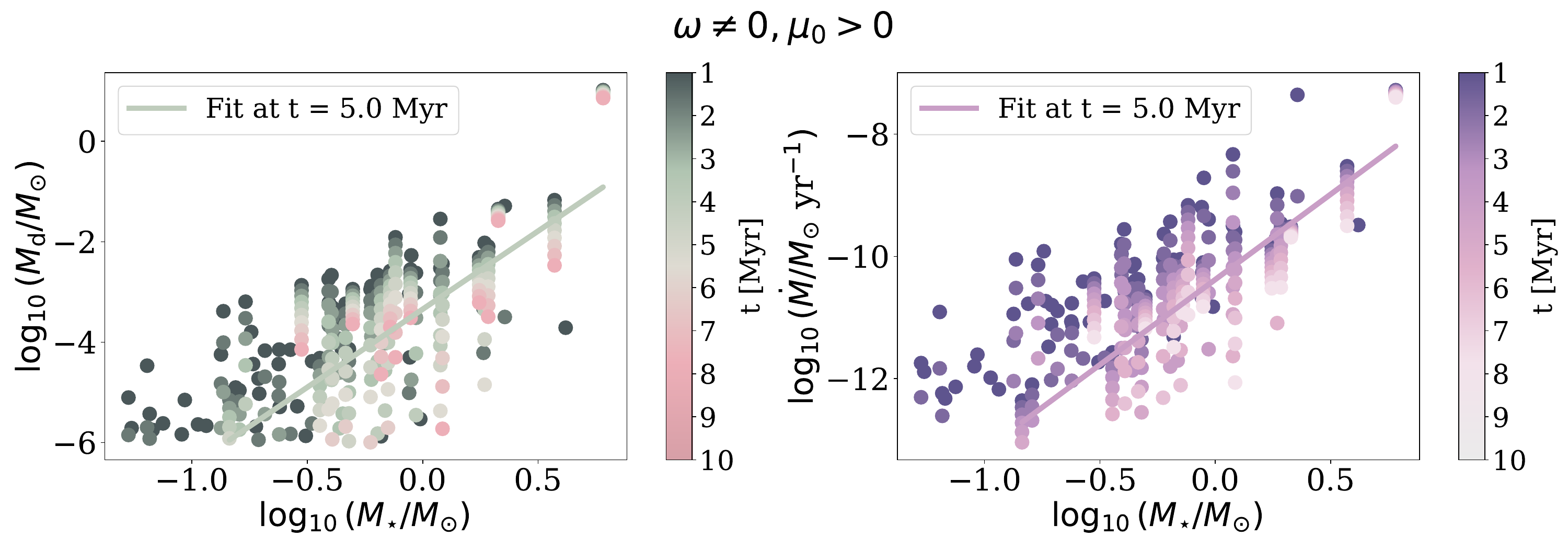}
    \caption{Same as Figure \ref{fig:brokencorr_MHD} with the addition of a spread in the initial correlations between the disc properties and the stellar mass ($\sigma_{M_{\mathrm{d}}}(0) = 0.65$ dex, $\sigma_{R_{\mathrm{d}}}(0) = 0.52$ dex). Despite the linear correlation being readily broken in theory (see Section \ref{subsubsec:calculations_MHD_omeganot0}), the scatter introduced by the spread in the initial conditions simulates the correlation also at more evolved ages. As an example, we show the fitted line at 5 Myr (the age of the oldest observed population) in both panels ($\log_{10}(M_{\mathrm{d}}/M_{\odot}) = 3.1 \log_{10}(M_{\star}/M_{\odot}) - 3.3$, $\log_{10}(\dot M/M_{\odot} \mathrm{yr}^{-1}) = 2.5 \log_{10}(M_{\star}/M_{\odot}) - 10.4$).}
    \label{fig:mhd_omeganotzero}
\end{figure*}

The introduction of an observationally-motivated spread in the initial conditions is crucial to produce realistic population synthesis model. In the purely viscous case, \cite{Somigliana+2022} have shown how the spread does not significantly impact the evolution of either the $M_{\mathrm{d}} - M_{\star}$ or $\dot M - M_{\star}$ correlations; the shape of the curves (grey in Figure \ref{fig:slopes_mhd_omegazero}) is unaffected, except for their starting point, and the statistical fluctuation - determined as the interval between the 25th and 75th percentile out of 100 realisations of numerical simulations with the same initial conditions - is of the order $\sim 0.1$ for both slopes. This is a factor two less than the smallest observational uncertainty, and therefore does not produce a detectable difference in the predicted results.

Following \cite{Somigliana+2022}, we set $\sigma_{M_{\mathrm{d}}}(0) = 0.65$ dex and $\sigma_{R}(0) = 0.52$ dex (determined from \citealt{Ansdell+2017} and \citealt{Testi+2022-Ofiucone}) for the log-normal distributions of $M_{\mathrm{d}}(0)$ and $R_{\mathrm{d}}(0)$ in the hybrid scenario (with $\alpha_{\mathrm{SS}} = \alpha_{\mathrm{DW}} = 10^{-3}$ hence $\psi =1$, $\lambda = 3$, $\omega = 0$). We find that, just like in the purely viscous case, a spread in the initial conditions only shifts the starting point of the curves (coloured in Figure \ref{fig:slopes_mhd_omegazero}) and does not have any significant effect on the shape of the evolution of the slopes (see Figure 5-6-7 in \citealt{Somigliana+2022}). The statistical fluctuation for both slopes is again of the order of 0.1, and therefore below the observational error and not impacting our predictions.

On the other hand, wind-driven models with increasing $\alpha_{\mathrm{DW}}$ in time and a spread in the initial conditions (Figure \ref{fig:mhd_omeganotzero}) behave quite differently from the theoretical expectation discussed in Section \ref{subsubsec:calculations_MHD_omeganot0}. As the spread introduces a stochastic component, the discs will have higher or lower masses and accretion rates with equal probability; the practical result for the initial correlations is that the bending towards lower stellar masses (approximately $\log_{10}(M_{\star}/M_{\odot}) < 0.5$ in Figure \ref{fig:brokencorr_MHD}) is lost to the stochastic displacement of the discs in the $M_{\mathrm{d}} - M_{\star}$ and $\dot M - M_{\star}$ planes. Actually, the resulting distribution of discs both in the $M_{\mathrm{d}} - M_{\star}$ and $\dot M - M_{\star}$ log-log plane does simulate a linear correlation; this implies that, while the stellar mass and the disc properties \textit{should not} exhibit a linear correlation already after a few $t_{\mathrm{acc}}$, the presence of a spread mimics such correlation, making the wind-driven scenario indistinguishable from the viscous and hybrid ones. The one feature that remains observable, despite the presence of a spread, is the removal of discs around more or less massive stars - depending on the value of $\mu_0$, as discussed in Section \ref{subsubsec:calculations_MHD_omeganot0}; in the simulation shown in Figure \ref{fig:mhd_omeganotzero} we have set $\mu_0 > 0$, implying that discs around less massive stars evolve more rapidly and are therefore more readily dispersed, as is visualised by the lack of sources around lower stellar masses at evolved ages. 

Summarising, an initial power-law correlation between the disc properties and the stellar mass would keep its power-law shape under wind-driven evolution, like in the viscous or hybrid case; however, the interpretation of the observed correlations is different depending on the theoretical framework. While the viscous and hybrid models \textit{preserve} an initially established correlation, making the characteristic evolution of the slopes $\lambda_{\mathrm{m}}$ and $\lambda_{\mathrm{acc}}$ a tracer of disc evolution itself, the apparent correlation observed in wind-driven populations is merely a signature of the initial spread, rather than the evolutionary mechanism at play.

\begin{figure*}
    \centering
    \includegraphics[width = \textwidth]{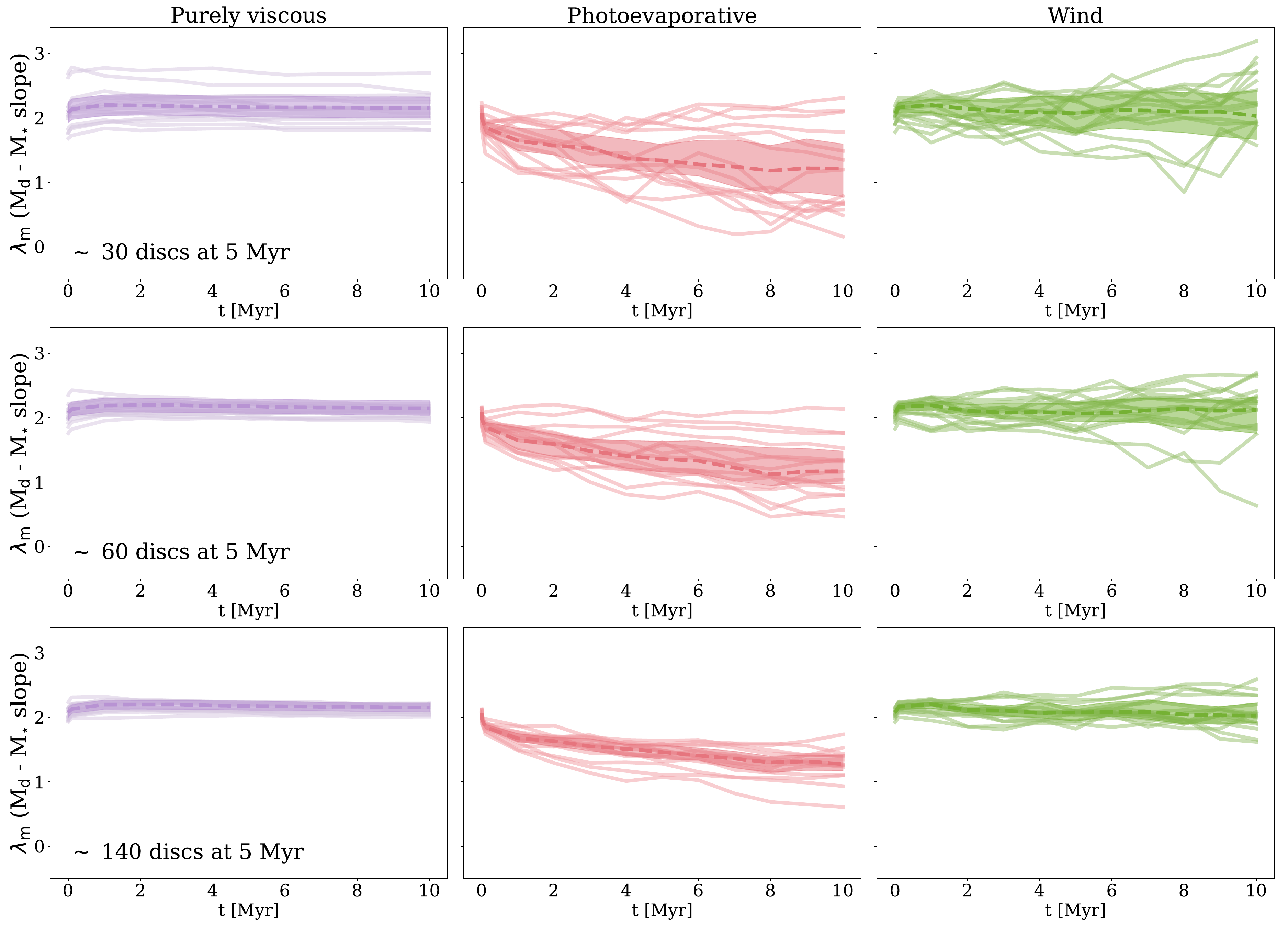}
    \caption{Time evolution of the slope of the $M_{\mathrm{d}} - M_{\star}$ correlation for 15 statistical realisations in the purely viscous (left, $\alpha_{\mathrm{SS}} = 10^{-3}$), photoevaporative (centre, $\dot M_w = 4 \times 10^{-10} \mathrm{M}_{\odot}/\mathrm{yr}$, $\alpha_{\mathrm{SS}} = 10^{-3}$) and wind-driven (right, $\omega = 0.25$, $\lambda = 3$, $\alpha_{\mathrm{DW}} = 10^{-3}$) model. The dashed lines show the median evolution, while the shaded area represents the interval between the 25th and the 75th percentiles. The three rows show different sample sizes, increasing from top to bottom. The initial size of each population was chosen to obtain a certain number of discs at 5 Myr (the age of the oldest observed population, Upper Sco) with the different disc fractions. In the top row, we match the current size of the Upper Sco sample($\sim 30$ objects, \citealt{Testi+2022-Ofiucone}) with both accretion rate and disc mass measurements; the middle row shows double the current sample size ($\sim 60$ objects), while the bottom row assumes a complete sample ($\sim 140$ objects). While the viscous model produces a remarkably similar evolution for all simulations, the latter two show stochastic oscillations from one realisation to another, suggesting that disc dispersal impacts the observed slope more than the evolutionary model does - at least with the currently available sample sizes; increasing the number of sources significantly mitigates the oscillations. The slope of the $\dot M - M_{\star}$ correlation behaves the same way.}
    \label{fig:stochastic_evolution}
\end{figure*}

\subsection{Accounting for disc dispersal: internal photoevaporation} \label{subsec:photoev}

Out of the three theoretical scenarios discussed so far, wind-driven evolution is the only one that manages to reproduce the disc and accretion fraction (as measured by \citealt{Hernandez+2007} and \citealt{Fedele+2010} respectively) - and therefore, the only one whose predictions can reasonably be compared with observations. Traditionally, the problem of disc dispersal in viscous populations is addressed by including internal photoevaporation (see e.g., \citealt{Hollenbach1994, ClarkeGendrinSoto2001, AlexanderClarkePringle2006a, AlexanderClarkePringle2006b}): in this Section, we discuss the impact of internal photoevaporation on the previously described expectations for the evolution of the slopes in the purely viscous scenario. As the lack of analytical solutions to the general equation \eqref{eq:master_equation} does not allow for analytical arguments, we base the following discussion on physical considerations.

Internal photoevaporation is a threshold process, that kicks in after the accretion rate drops below the photoevaporative mass-loss rate \citep{ClarkeGendrinSoto2001}. The moment where the effect of photoevaporation becomes non-negligible depends therefore on the initial accretion rate: assuming for simplicity a fixed photoevaporation rate for the whole population, as the accretion rate scales positively with the stellar mass we can expect discs around lower mass stars to show the effects of photoevaporation earlier. Moreover, given that the disc mass also scales positively with the stellar mass, said sources correspond to the less massive ones in the population. From these considerations, we can expect discs with lower initial mass to be the first ones to be affected by photoevaporation, causing a steepening of the $M_{\mathrm{d}} - M_{\star}$ and $\dot M - M_{\star}$ correlations. However, photoevaporation also disperses discs: removing sources from the population may alter the expected behaviour, hence the need to perform numerical simulations to understand the evolution of the correlations for a population of discs undergoing internal photoevaporation. Our simulations remove discs either because the photoevaporative gap becomes too large, or because the disc mass or accretion rate fall below a certain detectability threshold. As we mentioned above, the dispersal timescale $t_{\mathrm{disp}}$ might also play a role, if it has a different scaling with the stellar mass with respect to the accretion timescale ($t_{\nu}$ in the viscous case). Our numerical implementation of internal photoevaporation follows \cite{Owen+2010}, and we assume a mass-loss rate of $4 \times 10^{-10}$ M$_{\odot}$ yr$^{-1}$ - which allows us to reproduce the observed disc fraction for the set of parameters of our simulation - for all discs in the population; therefore, no further $M_{\star}$ dependence is introduced, and the $t_{\nu} - M_{\star}$ scaling is the only one that matters. We stress that a stellar mass-dependent photoevaporative rate \textit{is} expected \citep{Picogna+2021}: we explore the influence of such dependence on disc observables in an upcoming work (Malanga et al. in prep.). We discuss the results of our \texttt{Diskpop} simulations in the following Section.

\subsection{What are the slopes tracing?}\label{subsec:what_do_they_trace}

\begin{figure*}
    \centering
    \includegraphics[width = \textwidth]{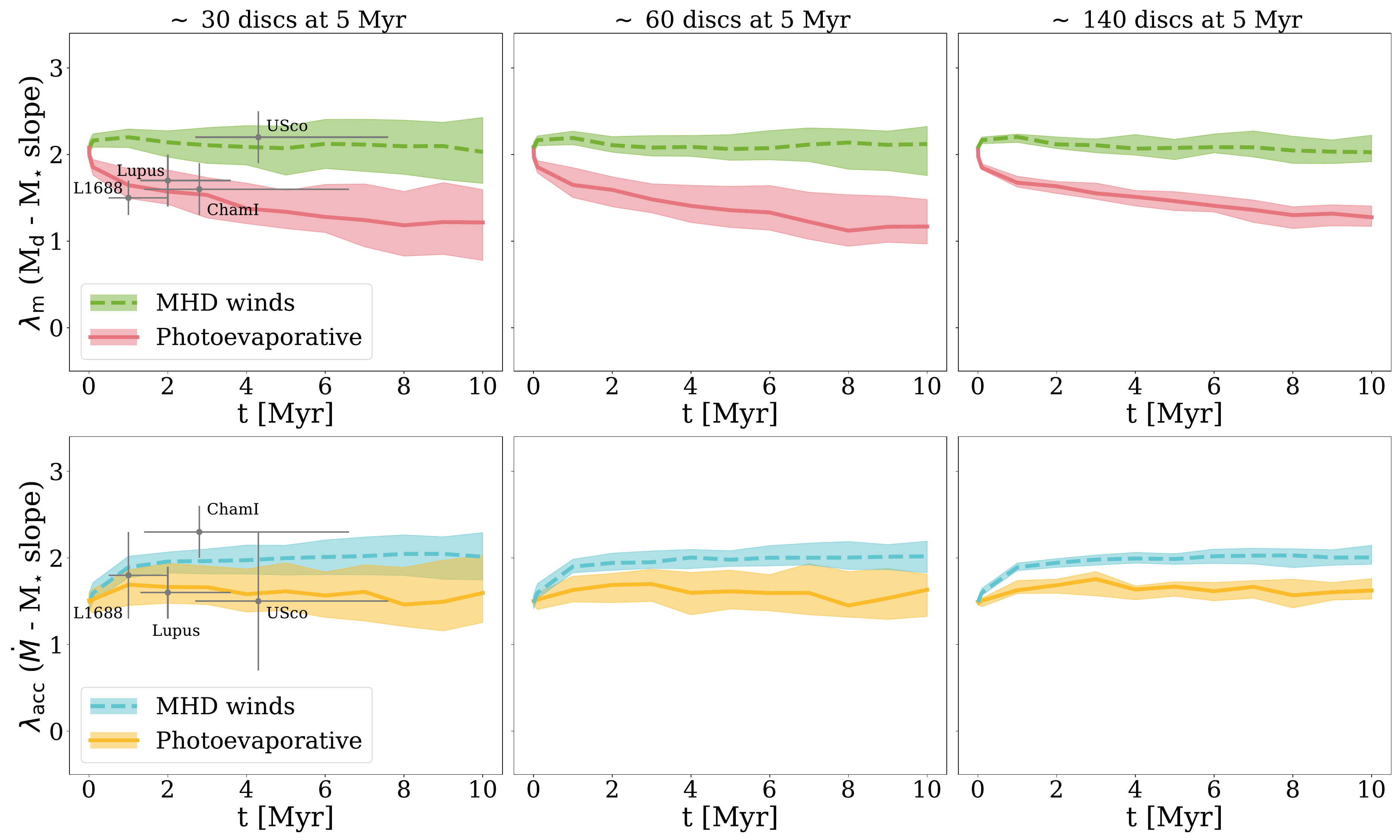}
    \caption{Comparison of the time evolution of $\lambda_{\mathrm{m}}$ (top) and $\lambda_{\mathrm{acc}}$ (bottom) between the viscous+photoevaporative and wind-driven case including measured slopes from four star-forming regions. To account for statistical fluctuations, each simulation combines 100 realisations of the same initial conditions: the lines show the median evolution, while the shaded area represent the interval between the 25th and 75th percentiles. The simulations in the three columns differ for the initial number of discs, determined to obtain a specific sample size at 5 Myr - currently available sample (left), double the currently available sample (centre) and the complete sample (right). Observed slopes from \cite{Testi+2022-Ofiucone}.}
    \label{fig:comparing_models}
\end{figure*}

Figure \ref{fig:stochastic_evolution} shows the time evolution of the $M_{\mathrm{d}} - M_{\star}$ slope for 15 realisations with the same initial conditions for the viscous plus photoevaporative (central panel) and wind-driven (right panel) models, both reproducing the observed disc and accretion fractions, compared to the purely viscous scenario (left panel). The initial number of discs in the populations was determined to recover a certain number of objects at 5 Myr (increasing from top to bottom), and varies in the different simulations, as the decline (or lack thereof) of the disc fraction is model-dependent (see Figure 6 of \citealt{Somigliana+2023}). 

The number of objects in the simulation displayed in the first row was set to obtain $\sim$ 30 discs at 5 Myr, corresponding to the currently available sample size in Upper Sco, the oldest observed star-forming region. In the left panel, we see how the evolution of the slope in the purely viscous model is not significantly affected by a spread in the initial conditions: the single realisations resemble each other remarkably well, the only difference being the starting point of the curve (as found by \citealt{Somigliana+2022}). On the other hand, the photoevaporative and wind-driven models have a dissimilar behaviour: each realisation can deviate substantially from the others, as we can particularly notice by the location, amplitude and direction of the bumps. The key difference between these models and the purely viscous case is disc dispersal: the stochastic nature of the slope evolution suggests that it does not trace the underlying secular disc evolution, like in the viscous scenario, but rather carry the signatures of disc dispersal itself - making it impossible to use the evolution of the slopes as a proxy for accretion mechanisms. There are two main factor that play a role in this context:

\begin{enumerate}
    \item \textbf{Initial conditions and spread in the correlations.} The exact evolution of the slopes will depend on the initial conditions, both of the disc mass-stellar mass correlations themselves and of the population-wide parameters. Furthermore, the removal of discs from the population would not impact the results of the fitting procedure only if there was perfect correlations between the disc parameters and the stellar mass; with a spread in the initial correlations, on the other hand, the results may differ depending on \textit{which} discs in the population are dispersed;
    
    \item \textbf{Small number statistics.} Depending on the initial number of objects, disc removal can lead to small samples - so small that it might lead to low number statistics issues. This is the case for the top row of Figure \ref{fig:stochastic_evolution}, where the number of objects at 10 Myr is of the order of 10 or lower.
\end{enumerate}

In this work, we have used one specific set of parameters (summarised in Table \ref{tab:params}), determined following \cite{Somigliana+2022} (viscous model) and \cite{Tabone+2022MHDTheory} (hybrid and wind-driven), and we leave a deeper exploration of the parameters space to a future work. While the exact shape of the slope evolution, and therefore the accretion model signature, might depend on the initial conditions, the top panel of Figure \ref{fig:stochastic_evolution} shows that with the current sample sizes the noise dominates over the physical evolution. However, the currently available sample of the oldest star-forming region (Upper Sco) with both disc masses (derived from the millimetre flux, \citealt{Barenfeld2016-masssurvey}) and accretion rate \citep{Manara+2020-UpperSco} estimates is highly incomplete; it is therefore worth investigating whether a higher level of completeness would help reducing the entity of the oscillations, allowing to disentangle between the different evolutionary models.

The central and bottom rows of Figure \ref{fig:stochastic_evolution} show how a larger sample would impact the oscillations of the slope evolution. The simulations in the middle row are performed imposing a double sample size at 5 Myr with respect to the current one ($\sim 60$ discs), while in the bottom row we assume to have the complete Upper Sco sample, totalling $\sim 140$ discs (Carpenter et al. in prep.). We remind that we focus on Upper Sco as the oldest observed star-forming region, which makes it the most affected by disc dispersal.

As expected, statistical significance increases with a larger sample, leading to a decreased impact of the oscillations on the global slope evolution; with the complete sample, in particular, we can reduce the spread in the evolution by a factor of $\sim 2$ compared to the current available data. This argument confirms the importance of larger sample sizes in discriminating between the viscous and wind-driven models, as already suggested by \cite{Alexander+2023} in the context of the accretion rates distribution.

As we mentioned in Section \ref{subsec:photoev}, our simulations consider discs as dispersed if their masses or accretion rates fall below the imposed detectability threshold of $10^{-12}$ M$_{\odot}$ yr$^{-1}$. We have also included a threshold in accretion rates in post-processing to account for non-accreting objects. From the observational point of view, this latter selection depends on both the instrumental sensitivity and the definition of disc itself: how Class II objects are defined, and in turn how Class III sources are removed from the observed samples, impacts the resulting slope. Summarising, with the current sample sizes, the evolution of the slopes is significantly more affected by disc dispersal than it is by secular evolution; therefore, at the state of the art, it cannot be used as a proxy to disentangle between the different evolutionary models. Increasing the sample size would allow to reduce the effect of low number statistics, potentially allowing to observe the different evolution of the slopes under the two evolutionary mechanisms; we further discuss this possibility in the following Section.


\begin{table*}[h]
\begin{minipage}{\textwidth} 
    \centering
    \vspace{0.3cm}

    \begin{tabular}[H]{|P{1cm} | P{1cm} P{2cm} | P{1cm} P{1.9cm} | P{1.6cm} P{5.2cm} | P{0.2cm}|}
        \multicolumn{8}{c}{ $\mu_0 = \delta_0 - \zeta_0 \xi$ }\\[5pt]
        \cline{1-7} \arrayrulecolor{red}\cline{8-8}\arrayrulecolor{black}
        $\xi$ & $\mu_0$ & $t_{\mathrm{acc}, 0} - M_{\star}$ & $\zeta_0$ & $R_{\mathrm{d}} - M_{\star}$ & $\delta_0$ & \multicolumn{1}{c!{\color{red}\vrule}}{$M_{\mathrm{d}}-M{\star}$ (a) and $\dot M-M{\star}$ (b)} & \multicolumn{1}{c!{\color{red}\vrule}}{$\Rightarrow$} \\[3pt]
        \hline
        
        \multirow{3}{*}{0} & \cellcolor{bg_visc_light}  $ < 0 $ & \cellcolor{bg_visc_light} $t_{\mathrm{acc}, 0} \downarrow M_{\star} \uparrow$ & \multicolumn{2}{c|}{\cellcolor{bg_visc_light} any} & \cellcolor{bg_visc_light} $ < 0 $ & \multicolumn{1}{c!{\color{red}\vrule}}{(a) shallower than (b) \cellcolor{bg_visc_light}} & \multicolumn{1}{c!{\color{red}\vrule}}{$=$ \cellcolor{bg_visc_light}} \\
         
         & \cellcolor{bg_visc} $0$ & \cellcolor{bg_visc} $t_{\mathrm{acc}, 0} \leftrightarrow M_{\star} \uparrow$ & \multicolumn{2}{c|}{\cellcolor{bg_visc} any} & \cellcolor{bg_visc} 0 & \multicolumn{1}{c!{\color{red}\vrule}}{(a) as steep as (b) \cellcolor{bg_visc}} &  \multicolumn{1}{c!{\color{red}\vrule}}{$=$ \cellcolor{bg_visc}} \\
         
         & \cellcolor{bg_visc_light} $ > 0 $ & \cellcolor{bg_visc_light} $t_{\mathrm{acc}, 0} \uparrow M_{\star} \uparrow$ & \multicolumn{2}{c|}{\cellcolor{bg_visc_light} any} & \cellcolor{bg_visc_light} $ > 0 $ & \multicolumn{1}{c!{\color{red}\vrule}}{(a) steeper than (b)\cellcolor{bg_visc_light}} & \multicolumn{1}{c!{\color{red}\vrule}}{$=$ \cellcolor{bg_visc_light}} \\
        \hline


        
         & \cellcolor{bg_mhd_light} & \cellcolor{bg_mhd_light} & \cellcolor{bg_mhd_light} $ \leq 0 $ & \cellcolor{bg_mhd_light} $R_{\mathrm{d}} \downarrow \leftrightarrow M_{\star} \uparrow$ & \cellcolor{bg_mhd_light} $ \leq 0 $ & \multicolumn{1}{c!{\color{red}\vrule}}{ (a) shallower or as steep as (b) \cellcolor{bg_mhd_light}} & \multicolumn{1}{c!{\color{red}\vrule}}{ $=$ \cellcolor{bg_mhd_light}}\\ 

        & \cellcolor{bg_mhd_light} \multirow{-2}{*}{$ < 0 $} & \cellcolor{bg_mhd_light} \multirow{-2}{*}{$t_{\mathrm{acc}, 0} \downarrow M_{\star} \uparrow$} & \cellcolor{bg_mhd_light} $ > 0 $ & \cellcolor{bg_mhd_light} $R_{\mathrm{d}} \uparrow M_{\star} \uparrow$ & \cellcolor{bg_mhd_light} $ (- \infty, \zeta_0 \xi) $ & \multicolumn{1}{c!{\color{red}\vrule}}{(a) shallower or steeper than (b)\footnote{\label{fn:table_1} $\zeta_0 \xi $ is positive, therefore $\delta_0$ can either be negative (implying (a) shallower than (b)) or positive (implying (a) steeper than (b)).} \cellcolor{bg_mhd_light}} & \multicolumn{1}{c!{\color{red}\vrule}}{$\neq$ \cellcolor{bg_mhd_light}}\\


        & \cellcolor{bg_mhd} & \cellcolor{bg_mhd} & \cellcolor{bg_mhd} $ \leq 0 $ & \cellcolor{bg_mhd} $R_{\mathrm{d}} \downarrow \leftrightarrow M_{\star} \uparrow$ & \cellcolor{bg_mhd} $ \leq 0 $ & \multicolumn{1}{c!{\color{red}\vrule}}{ (a) shallower or as steep as (b) \cellcolor{bg_mhd}} & \multicolumn{1}{c!{\color{red}\vrule}}{$\circ$ \cellcolor{bg_mhd}} \\ 
         
        & \cellcolor{bg_mhd} \multirow{-2}{*}{$ 0 $} & \cellcolor{bg_mhd} \multirow{-2}{*}{$t_{\mathrm{acc}, 0} \leftrightarrow M_{\star} \uparrow$} & \cellcolor{bg_mhd} $ > 0 $ & \cellcolor{bg_mhd} $R_{\mathrm{d}} \uparrow M_{\star} \uparrow$ & \cellcolor{bg_mhd} $ \geq 0 $ & \multicolumn{1}{c!{\color{red}\vrule}}{(a) steeper or as steep as (b) \cellcolor{bg_mhd}} &  \multicolumn{1}{c!{\color{red}\vrule}}{$=$ \cellcolor{bg_mhd}} \\

         
         & \cellcolor{bg_mhd_light} & \cellcolor{bg_mhd_light} & \cellcolor{bg_mhd_light} $ \leq 0 $ & \cellcolor{bg_mhd_light} $R_{\mathrm{d}} \downarrow \leftrightarrow M_{\star} \uparrow$ & \cellcolor{bg_mhd_light} $\left(\zeta_0 \xi, +\infty \right)$ & \multicolumn{1}{c!{\color{red}\vrule}}{(a) shallower or steeper than (b)\footnote{\label{fn:table_1.2} $\zeta_0 \xi $ is negative, therefore the same argument as in \ref{fn:table_1} holds.} \cellcolor{bg_mhd_light}} & \multicolumn{1}{c!{\color{red}\vrule}}{ $\circ$ \cellcolor{bg_mhd_light}} \\
         
         \multirow{-6}{*}{(0, 1)} & \cellcolor{bg_mhd_light} \multirow{-2}{*}{$ > 0 $} & \cellcolor{bg_mhd_light} \multirow{-2}{*}{$t_{\mathrm{acc}, 0} \uparrow M_{\star} \uparrow$} & \cellcolor{bg_mhd_light} $ > 0 $ & \cellcolor{bg_mhd_light} $R_{\mathrm{d}} \uparrow M_{\star} \uparrow$ & \cellcolor{bg_mhd_light} $\left(\zeta_0 \xi, +\infty \right)$ & \multicolumn{1}{c!{\color{red}\vrule}}{(a) steeper than (b) \cellcolor{bg_mhd_light}} & \multicolumn{1}{c!{\color{red}\vrule}}{$=$ \cellcolor{bg_mhd_light}} \\ 
 
        \cline{1-7} \arrayrulecolor{red}\cline{8-8}\arrayrulecolor{black}

    \end{tabular}

        \vspace{0.4cm}

    \begin{tabular}[H]{|P{1.6cm} P{5cm}|}
        \multicolumn{2}{c}{ $\delta_{\mathrm{evo}} = \xi ( \zeta_0 - \mu_0), \quad \xi \neq 0$ }\\[5pt]
        \hline
        $\delta_{\mathrm{evo}}$ & $M_{\mathrm{d}}-M{\star}$ (a) and $\dot M-M{\star}$ (b) \\[3pt]
        \hline

        
         < 0 &  (a) shallower than (b) \\

        
        \cellcolor{bg_2} 0 & \cellcolor{bg_2} (a) as steep as (b) \\

    
         > 0 & (a) steeper than (b) \\

         \hline

    \end{tabular}
\end{minipage}

\vspace{0.2cm}

    \caption{Summary of the different possible theoretical scenarios described in Section \ref{sec:time_evo_analytical} to visualise the relative signs of the parameters at play. From left to right in the top table, the columns show (i) $\xi$, a proxy for the evolutionary model (viscous if $\xi = 0$, hybrid or wind-driven otherwise); (ii) $\mu_0$, the slope of the $t_{\mathrm{acc}, 0} - M_{\star}$ correlation and its implication on the correlation itself; (iii) $\zeta_0$, the slope of the $R_{\mathrm{d}}-M_{\star}$ correlation, and its implication on the correlation itself; (iv) $\delta_0 = \lambda_{\mathrm{m}, 0} - \lambda_{\mathrm{acc}, 0}$, the difference between the initial slopes of the $M_{\mathrm{d}}-M_{\star}$ and $\dot M - M_{\star}$ correlations and its implication on their relative steepness. The final column summarises whether the signs of $\mu_0$ and $\delta_0$ are \textit{necessarily the same} ($=$), \textit{necessarily opposite} ($\neq$) or \textit{can be either} ($\circ$). When discussing the implications on correlations, up(down)wards arrows represent an in(de)crease of the parameters, while horizontal arrows describes the lack of correlation. The different cell colours are purely meant to guide the eye. The top table links the initial conditions, while the bottom table summarises the implications of the evolved difference in the slopes.}
    \label{tab:table_results}

\end{table*}


\section{The observational relevance of the slopes}\label{sec:discussion}

Observed star-forming regions have both a spread in the initial conditions in addition to some disc dispersal mechanism (be it photoevaporation or MHD winds); as we discussed in Section \ref{subsec:what_do_they_trace}, with the current sample sizes, the statistical significance of the observationally-determined slopes is undermined and their evolution traces disc dispersal, rather than the accretion mechanism. In this Section, we perform a statistical analysis of our simulated slopes and compare them with the currently available measurements; furthermore, we show the relevance of measuring the slopes despite these limitations and discuss the conditions under which they allow us to put constraints on disc evolution.

\subsection{Comparison of different evolutionary models} \label{subsec:model_comparison}

Figure \ref{fig:comparing_models} shows the comparison of the evolution of the slopes between the viscous + photoevaporative (solid line) and wind-driven (dashed line) models for both the $M_{\mathrm{d}}-M_{\star}$ and $\dot M - M_{\star}$ correlations (top and bottom row, respectively), including the measured slopes in four star-forming regions from \cite{Testi+2022-Ofiucone} as grey dots. As we mentioned above, both models lead to disc dispersal consistently with the observed disc and accretion fraction (shown in Figure 6 of \citealt{Somigliana+2023}); the three columns show simulations performed with a different initial number of discs, increasing from left to right, to obtain a different sample size at 5 Myr according to the predicted decline of observed discs. Like in Figure \ref{fig:stochastic_evolution}, the number of objects at 5 Myr is $\sim 30$, $\sim 60$ and $\sim 140$ from left to right, increasing from the currently available measurements in Upper Sco to the virtually complete sample. To estimate the effect of statistical fluctuations, given by the spread in the initial conditions, we ran 100 simulations for each set up: the solid and dashed lines represent the median evolution, while the intervals between the 25th and 75th percentile are visualised by the shaded areas. The growing shaded area, particularly visible with smaller sample sizes, is representative of the decreasing amount of sources on which the fit is performed: with the current sample size (left column), which leads to $\sim 30$ discs at 5 Myr, we end up with a 1$\sigma$ deviation from the median value of $\sim 0.5 - 0.6$. Larger sample sizes significantly reduce the scatter, leading to $\sigma \sim 0.4$ with a double sample and $\sigma \sim 0.2$ for the complete sample, reducing the current one by a factor 3. As mentioned in Section \ref{subsec:photoev}, with the currently available number of objects the dominant role in the evolution of the slopes is played by disc dispersal, which makes it difficult to trace the imprint of the secular evolution. The expected Upper Sco complete sample (right column) allows for a better separation between the two models - particularly for the $M_{\mathrm{d}}-M_{\star}$ correlation: the expected slope in the two scenarios differs by $\sim 0.5$, while the typical uncertainty of the currently measured slopes is between 0.2 and 0.3. Larger sample sizes would further decrease this uncertainty, allowing us to discriminate between the two models based on the slope evolution. 

The observed slopes (from \citealt{Testi+2022-Ofiucone}) are only included in the left column of Figure \ref{fig:comparing_models} as they refer to the current sample size. The main source of uncertainty in the current measurements is given by Upper Sco, mainly due to the incomplete sample; moreover, it is worth pointing out that external photoevaporation is likely to play a significant role in this region (Anania et al. in prep.). This comparison is meant as a first glance of the parameters space of the observed slopes, and we anticipate a proper exploration of the initial conditions once the full sample will be available.

In the following Section, we discuss the other constraints that we can put on disc evolution, besides identifying the driving accretion mechanism.

\subsection{What are the slopes telling us, then?}\label{subsec:what_do_they_tell_us}

Despite not allowing to conclusively discriminate between different evolutionary scenarios with the current sample sizes, the slopes of the $M_{\mathrm{d}} - M_{\star}$ and $\dot M - M_{\star}$ correlations can still help with constraining other properties from the theoretical considerations presented in Section \ref{sec:time_evo_analytical}, which we summarise in Table \ref{tab:table_results}. If we \textit{assume} an evolutionary model to begin with, and we can estimate (directly or indirectly) either $\mu_0$ or $\delta_0$, we can constrain the other parameter. When discussing the observational determination of what we have so far referred to as \textit{initial conditions}, it is important to clarify the meaning of "initial". \texttt{Diskpop} deals with and evolves Class II, potentially Class III, objects; hence, the initial conditions we input refer to the beginning of the Class II phase, where the protostellar collapse is over and the disc is already formed. From the observational point of view, this means that we expect $\delta_0$ and $\mu_0$ to refer to young Class II objects - around, or younger than, approximately 1 Myr. Earlier phases like the Class 0 and I need a dedicated study, as the accretion of the protostellar envelope is expected to play a prominent role in those stages.

In the following, we discuss the constraints we can put in both directions and comment on their feasibility based on the currently available estimates of $\mu_0$ and $\delta_0$.

\subsubsection{Constraining \texorpdfstring{$\delta_0$}{delta0} from \texorpdfstring{$\mu_0$}{mu0}}\label{subsec:constraining_delta_from_mu}

\cite{Ansdell+2017} claimed $\lambda_{\mathrm{m}}$ to be increasing with time. As shown by \cite{Somigliana+2022} and discussed in Section \ref{sec:time_evo_analytical}, increasing slopes imply that discs around less massive stars evolve faster than those around more massive stars; this can be interpreted in terms of increasing accretion timescale with stellar mass, which corresponds to $\mu_0 > 0$ (with $t_{\mathrm{acc}, 0} \propto {M_{\star}}^{\mu_0}$). In this section we discuss the implications of the increasing slope scenario on the initial conditions, $\lambda_{\mathrm{m}, 0}$ and $\lambda_{\mathrm{acc}, 0}$.

The top panel of Table \ref{tab:table_results} shows the relation between $\mu_0$ and $\delta_0$ in the different evolutionary models. As $\mu_0 = \delta_0 - \zeta_0 \xi$, in the viscous case (corresponding to $\xi = 0$) we have $\mu_0 = \delta_0$ as mentioned in Section \ref{sec:time_evo_analytical}. This means that, to recover the suggested increasing slopes scenario, $\delta_0$ necessarily needs to be positive - regardless of the value of any other parameters: this translates to the initial $M_{\mathrm{d}} - M_{\star}$ correlation being steeper than $\dot M - M_{\star}$. In the hybrid and wind-driven case, instead, the implication is less straightforward as a positive $\mu_0$ can lead to opposite signs of $\delta_0$: this is determined by the scaling of the disc radius with the stellar mass, which is suggested to be (weakly) positive from observational evidences (e.g., \citealt{Andrews2018Sizes}). In principle, as $\delta_0 \in (\zeta_0 \xi, + \infty)$, the sign of $\zeta_0$ determines whether negative values of $\delta_0$ are possible; however, as $\xi$ is a small number (0.1 in this work), only a limited area of the parameters space would lead to a negative $\delta_0$. Summarising, if we assume increasing slopes ($\mu_0 > 0$) we can constrain the sign of $\delta_0$ regardless of the evolutionary model assumed: in both cases $\delta_0$ needs to be positive, which leads to an initially steeper $M_{\mathrm{d}} - M_{\star}$ than $\dot M - M_{\star}$ correlation.

\subsubsection{Constraining \texorpdfstring{$\mu_0$}{mu0} from \texorpdfstring{$\delta_{\mathrm{evo}}$}{deltaevo}}\label{subsec:constraining_mu_from_delta}

Instead of assuming increasing slopes, we can start from the currently measured values of $\delta_{\mathrm{evo}}$ and estimate $\delta_0$ in the different evolutionary models. In the viscous case, because $\delta_{\mathrm{evo}} = 0$, we focus on the value of the single slopes instead: as $\lambda_{\mathrm{m, evo}} = \lambda_{\mathrm{acc, evo}} = \delta_0/2 + \lambda_{\mathrm{m}, 0}$, the measured final value of the slopes does not help in constraining $\delta_0$ as it also depends on $\lambda_{\mathrm{m}, 0}$. In the hybrid case, instead, we have $\delta_{\mathrm{evo}} = \xi(\zeta_0 - \mu_0)$, meaning that if we can determine the sign of $\delta_{\mathrm{evo}}$ we can constrain that of $\mu_0$ as well. While in principle the sign of $\zeta_0$ influences that of $\mu_0$, as we mentioned above $\zeta_0$ is likely a small number: therefore, effectively, $\delta_{\mathrm{evo}}$ and $\mu_0$ have opposite signs for the vast majority of the parameters space.

Assuming that the observed disc populations can be considered evolved enough for the above arguments to hold, we can estimate $\delta_{\mathrm{evo}}$ from the most recent and homogeneous measurements available of $\lambda_{\mathrm{m}}$ and $\lambda_{\mathrm{acc}}$ \citep{Testi+2022-Ofiucone}. The resulting values of $\delta_{\mathrm{obs}}$ (which we label 'observed' as opposed to the theoretical expectation, 'evolved'), summarised in Table \ref{tab:delta_Testi}, are oscillating: out of the four regions L1668, Lupus, Chameleon I and Upper Sco, we find two positive and two negative median values. Moreover, in three cases out of four the uncertainties are so large that $\delta_{\mathrm{obs}}$ would be compatible with both a positive and a negative value. The difficulty in assessing the sign of $\delta_{\mathrm{obs}}$ from the current measurements of the slopes make constraining $\mu_0$ from $\delta_{\mathrm{evo}}$ not trivial. Larger sample sizes would give a better measurement of the slopes and reduce the uncertainty, leading to a more solid determination of the sign of $\delta_{\mathrm{obs}}$ - which would possibly allow to constrain $\mu_0$. 

Summarising, the (admittedly not robust) observational evidence pointing towards increasing accretion timescale with stellar mass allows us to constrain the initial correlations between the stellar mass and disc parameters; regardless of the evolutionary model considered, the initial slope of the $M_{\mathrm{d}}-M_{\star}$ correlation needs to be larger than that of $\dot M - M_{\star}$. The other way around, constraining the slope of the accretion timescale - stellar mass correlation from the difference between $\lambda_{\mathrm{m}}$ and $\lambda_{\mathrm{acc}}$ at the present time, requires sample sizes larger by at least a factor two.

\begin{table}[h]
    \centering
    \vspace{0.3cm}

    \begin{tabular}[H]{|P{1.5cm} | P{3.5cm} | P{1.5cm} |}

    \hline
    \hline

    Region & Median age [Myr] & $\delta_{\mathrm{obs}}$ \\

    \hline
    
    L1668 & 1 & $-0.3 \pm 0.5$ \\
    Lupus & 2 & $0.1 \pm 0.4$ \\
    Cha I & 2.8 & $-0.7 \pm 0.4$ \\
    Upper Sco & 4.3 & $0.7 \pm 0.8$ \\

    \hline

    \end{tabular}
    
    \hspace{1cm}
    
    \caption{Values of $\delta$ derived from the currently available measurements of $\lambda_{\mathrm{m}}$ and $\lambda_{\mathrm{acc}}$ \citep{Testi+2022-Ofiucone}.}
    \label{tab:delta_Testi}
\end{table}

\section{Conclusions}\label{sec:conclusions}

In this paper, we have investigated the impact of disc evolution models on the correlations between the stellar mass and the disc properties - especially the disc mass and the accretion rate. We have explored the purely viscous, wind-driven, viscous and wind hybrid, and photoevaporative models. Assuming power-law correlations to hold as initial conditions, $M_{\mathrm{d}}(0) \propto {M_{\star}}^{\lambda_{\mathrm{m}, 0}}$, $\dot M(0) \propto {M_{\star}}^{\lambda_{\mathrm{acc}, 0}}$, we performed analytical calculations (where possible) and population synthesis simulations for both evolutionary scenarios, and compared them with the purely viscous case discussed in \cite{Somigliana+2022}. Our main results are the following:

\begin{enumerate}

    \item The viscous and hybrid models change the slope of the initial correlations as function of the evolutionary time, but preserve their shape. In the wind-driven model, instead, the correlations deviate from the original power-law shape: this is visualised in the logarithmic plane as a bending of the linear correlation (see Figure \ref{fig:slopes_mhd_omegazero}). The bending direction is towards the less or more massive stars depending on the scaling of the accretion timescale with the stellar mass (positive and negative correlation respectively).
    
    \item The characteristic behaviour of the slopes in the wind-driven model is concealed by the presence of a spread in the initial conditions, which introduces a scatter in the correlations and makes it no longer possible to detect the bending (Figure \ref{fig:brokencorr_MHD}). This leads to a considerably similar evolution of the correlations in the different accretion models.

    \item Performing our simulations with evolutionary models that match the disc dispersal timescales (intrinsic in the wind-driven model and including internal photoevaporation in the viscous case), we find that the evolution of the slopes is significantly impacted by the removal of discs from the population (Figure \ref{fig:mhd_omeganotzero}). Different realisations of the same simulation dramatically differ from one another, and show a stochastic behaviour with large variations (Figure \ref{fig:stochastic_evolution}). This has both a physical (presence of a spread in the initial conditions) and a statistical (low number of objects left after a few Myr of evolution) reason.

    \item While a proper exploration of the parameters space, outside of the scope of this work, would be needed to assess the impact of the initial conditions, with the currently available sample sizes the noise dominates over the physical evolution.

    \item Increasing the sample size can mitigate the effects of disc dispersal on the evolution of the slope by removing the stochastic effects. We find that, for our parameters choice, the complete sample of Upper Sco ($\sim 140$ sources) at 5 Myr would reduce the oscillations enough to make the slopes a proxy for the evolutionary model (Figure \ref{fig:comparing_models}).

    \item While the currently available sample sizes do not yet allow to distinguish between the different evolutionary models, we can use them to put some constraints on the initial conditions. We find that in all evolutionary scenarios, the observational claim of increasing slopes leads to an initially steeper correlation between the disc mass and the stellar mass than between the accretion rate and the stellar mass. The other possible way, measuring the current slopes and inferring the correlation between the accretion timescale and the stellar mass from them, provides weaker constraints because of the high uncertainties in the current measurements.

    \item We have presented and released the 1D Python disc population synthesis code \texttt{Diskpop} and its output analysis library \texttt{popcorn}.
    
\end{enumerate}

In this work, we have shown how large enough samples of protoplanetary discs can provide a way of distinguishing between the evolutionary models (with a standard set of parameters) through the observation of the time evolution of the correlations between the disc properties and the stellar mass. We have shown how the stochastic fluctuations seen with the currently available observations could be significantly reduced if we had access to the complete Upper Sco sample, consisting of approximately 140 sources at 5 Myr. We strongly support the observational effort in the direction of obtaining larger amounts of data for evolved star-forming regions, and encourage the exploration of the parameters space beyond the standard case.

\begin{acknowledgements}

We thank an anonymous referee for their useful comments that helped us improve the clarity of the manuscript. This work was partly supported by the Italian Ministero dell’Istruzione, Universit\`{a} e Ricerca through the grant Progetti Premiali 2012-iALMA (CUP C52I13000140001), by the Deutsche Forschungsgemeinschaft (DFG, German Research Foundation) - Ref no. 325594231 FOR 2634/2 TE 1024/2-1, by the DFG Cluster of Excellence Origins (www.origins-cluster.de). This project has received funding from the European Union’s Horizon 2020 research and innovation program under the Marie Sklodowska- Curie grant agreement No 823823 (DUSTBUSTERS) and from the European Research Council (ERC) via the ERC Synergy Grant ECOGAL (grant 855130) and ERC Starting Grant DiscEvol (grant 101039651). GR acknowledges support from Fondazione Cariplo, grant No. 2022-1217. Views and opinions expressed are however those of the authors only and do not necessarily reflect those of the European Union or the European Research Council Executive Agency. Neither the European Union nor the granting authority can be held responsible for them.

\end{acknowledgements}

%
%

\bibliography{draft}
\bibliographystyle{aa}

\begin{appendix}

\section{Parameters used throughout the paper}\label{sec:appendix_parameters}

Table \ref{tab:params} shows the parameters used throughout the paper. The general simulation parameters refer to the initial correlations (distribution, slopes and spreads), while the following three tables are divided by evolutionary model and display the main parameters for each of them.

\begin{table}[h]
    \centering
    \vspace{0.1cm}

    General simulation parameters

    \vspace{0.1cm}
    \begin{tabular}[H]{P{3.5cm} | P{1.5cm} }

    \hline

    \rule{0pt}{1.5ex} $\lambda_{\mathrm{m}, 0}$ & 2.1 \\[0.1ex]
    \rule{0pt}{1.5ex}$\lambda_{\mathrm{acc}, 0}$ & 1.5 \\[0.1ex]
    \rule{0pt}{1.5ex}$\sigma_{\mathrm{M_d}}$ & $0.65$ dex \\[0.1ex]
    \rule{0pt}{1.5ex}$\sigma_{\mathrm{R_d}}$ & $0.52$ dex \\[0.1ex]
    \rule{0pt}{1.5ex}$M_{\mathrm{d}}$, $R_{\mathrm{d}}$ distribution & lognormal \\[0.1ex]

    \hline
    
    \end{tabular}

    \vspace{0.5cm}
    
    Viscous model

    \vspace{0.1cm}
    \begin{tabular}[H]{P{1.5cm} | P{1.5cm}}

    \hline

    \rule{0pt}{2.2ex} $\alpha_{\mathrm{SS}}$ & $10^{-3}$ \\[0.1ex]

    \hline

    \end{tabular}
    
    \vspace{0.5cm}

    Photoevaporative model

    \vspace{0.1cm}
    \begin{tabular}[H]{P{1.5cm} | P{3cm}}

    \hline

    \rule{0pt}{1.5ex} $\alpha_{\mathrm{SS}}$ & $10^{-3}$ \\[0.1ex]
    \rule{0pt}{1.5ex} $\dot M_{w}$ & $4 \times 10^{-10}$ M$_{\odot}$ yr$^{-1}$\\[0.1ex]

    \hline

    \end{tabular}

    \vspace{0.5cm}

    Wind-driven model

    \vspace{0.1cm}
    \begin{tabular}[H]{P{1.5cm} | P{1.5cm}}

    \hline

    \rule{0pt}{1.5ex} $\alpha_{\mathrm{SS}}$ & $10^{-3}$ \\[0.1ex]
    \rule{0pt}{1.5ex} $\alpha_{\mathrm{DW}}$ & $10^{-3}$ \\[0.1ex]
    \rule{0pt}{1.5ex} $\lambda$ & $3$ \\[0.1ex]
    \rule{0pt}{1.5ex} $\omega$ & $0.25$ \\[0.1ex]

    \hline

    \end{tabular}

    \hspace{1cm}
    
    \caption{Values of the parameters used throughout the paper, unless explicitly stated otherwise.}
    \label{tab:params}
\end{table}

\section{MHD model with \texorpdfstring{$\mu < 0$}{mu<0}}

\begin{figure}[h]
    \centering
    \includegraphics[width = 0.45\textwidth]{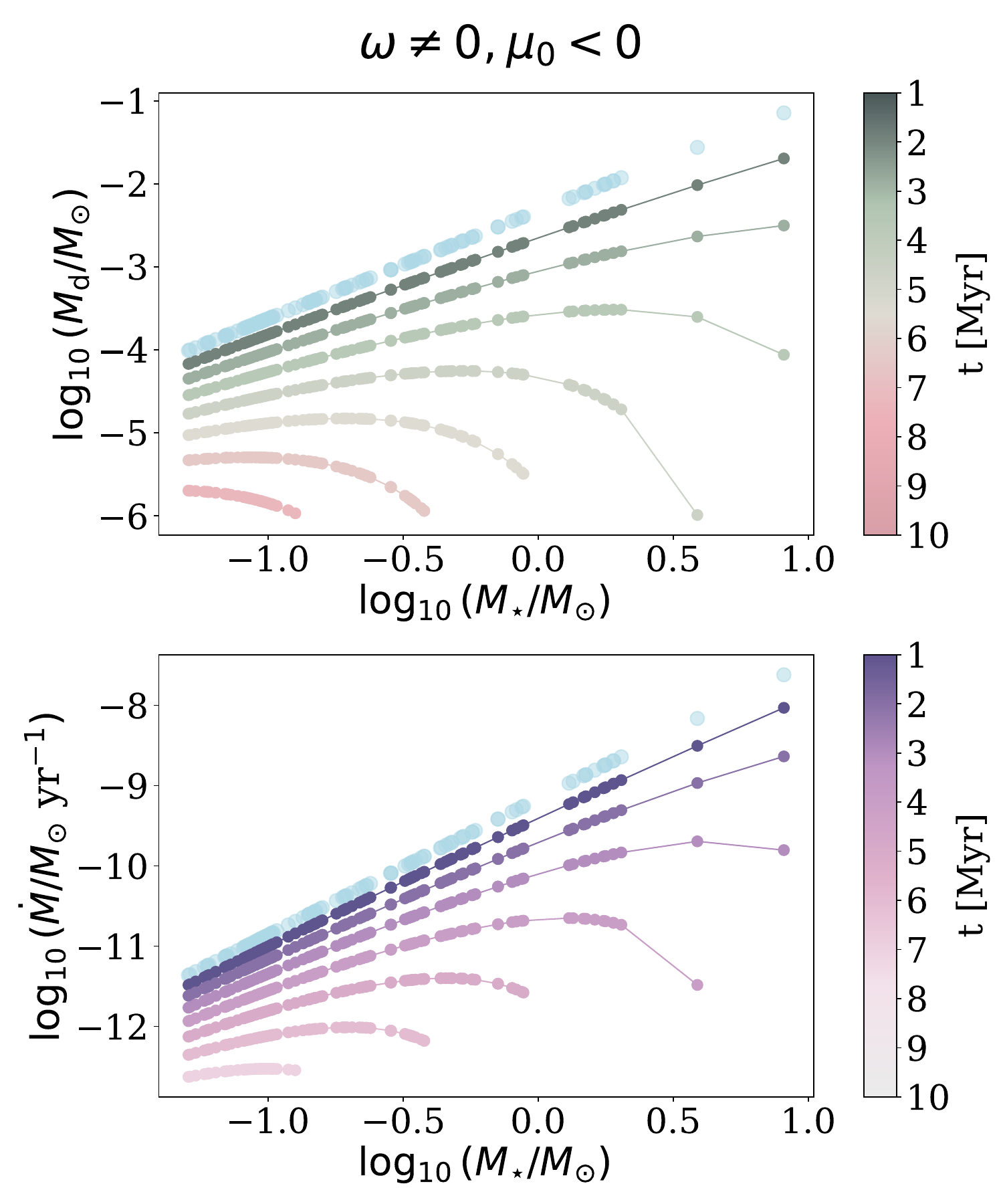}
    \caption{Same as Figure \ref{fig:brokencorr_MHD}, but with a choice of initial slopes resulting in a negative $\mu_0$ ($\lambda_{\mathrm{m}, 0} = 1.3, \lambda_{\mathrm{acc}, 0} = 1.7$). The bending of the linear correlation happens towards larger disc masses, in agreement with the prediction.}
    \label{fig:appendix_brokencorr_MHD}
\end{figure}

As mentioned in Section \ref{subsubsec:calculations_MHD_omeganot0}, the breaking of the linear correlation between the disc properties and the stellar mass happens towards higher or lower stellar masses depending on the value of $\mu_0$. Figure \ref{fig:brokencorr_MHD} shows the a simulation $\mu_0 > 0$, while in Figure \ref{fig:appendix_brokencorr_MHD} we show the opposite case. As $\mu_0$ is linked to $\delta_0$ through $\mu_0 = \delta_0 - \zeta_0 \xi$, if $\delta_0 < \zeta_0 \xi$ the resulting $\mu_0$ will be negative, leading to a specular bending of the correlation. Given that $\zeta_0 \xi$ is a small number ($\sim 0.1$ in our simulation), this generally requires a negative $\delta_0$. The simulation in Figure \ref{fig:appendix_brokencorr_MHD} has $\lambda_{\mathrm{m}, 0} = 1.3$ and $\lambda_{\mathrm{acc}, 0} = 1.7$, resulting in $\delta_0 = -0.4$.

\section{Validation of \texttt{Diskpop}}\label{sec:appendix_validation} 

\begin{figure*}
    \centering
    \includegraphics[width = \textwidth]{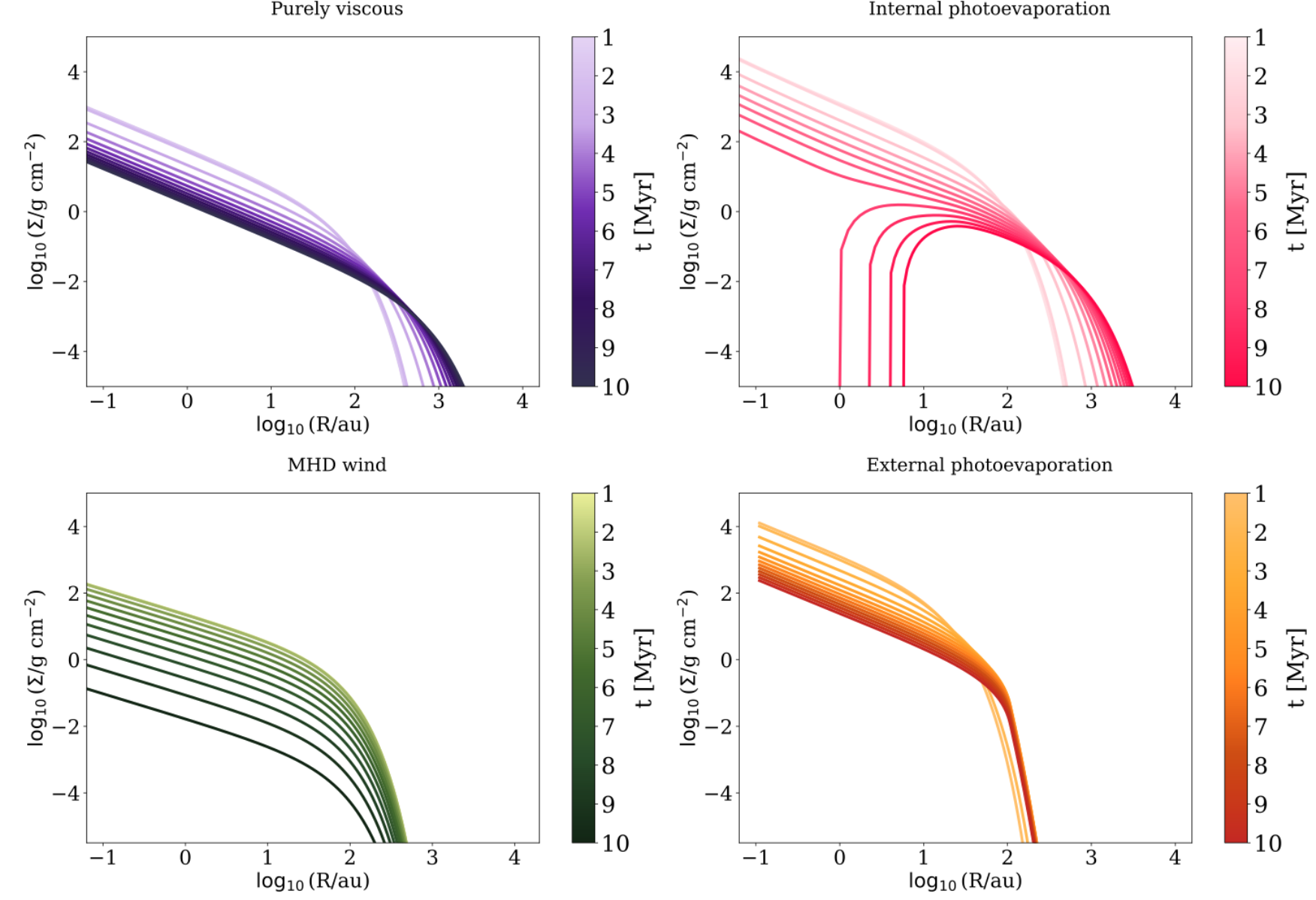}
    \caption{Gas surface density as a function of radius for a disc generated with Diskpop at different times as the colour bar shows. The four models are purely viscous (top left, $\alpha_{\mathrm{SS}} = 10^{-3}$), viscous including internal photoevaporation (top right, $\alpha_{\mathrm{SS}} = 10^{-3}$, $\dot M_{\mathrm{w}} = 4 \times 10^{-10}$ M$_{\odot} \mathrm{yr^{-1}}$), wind-driven (bottom left, $\alpha_{\mathrm{DW}} = 10^{-3}$, $\lambda = 3$, $\omega = 0.25$) and viscous including external photoevaporation (bottom right, $\alpha_{\mathrm{SS}} = 10^{-3}$, FUV = 100 $G_0$) respectively.}
    \label{fig:validation_sigma}
\end{figure*}

Figure \ref{fig:validation_sigma} shows the evolution of the gas surface density as a function of the disc radius in the cases of evolution driven by (i) viscosity, (ii) viscosity and internal photoevaporation, (iii) MHD winds, and (iv) viscosity and external photoevaporation for a single disc simulated with \texttt{Diskpop}. The top left panel, corresponding to viscous evolution (i), shows the key feature of viscous spreading: while the global surface density decreases as a consequence of the accretion onto the central star, the radial extent of the disc increases. This is a consequence of the redistribution of angular momentum, that causes part of the disc material to move towards larger radii. The top right panel, where the disc evolves under the combined effect of viscosity and internal photoevaporation (ii), shows the characteristic two-timescales behaviour \citep{ClarkeGendrinSoto2001}: the evolution is effectively viscous in the earliest stages, as long as the accretion rate is larger than the photoevaporative mass-loss rate; then, photoevaporation opens a cavity within the disc, which gets divided into an inner and an outer disc. The inner disc is less extended and therefore has a shorter viscous timescale, which means it evolves much faster and is quickly completely accreted onto the protostar; the outer disc instead keeps on evolving on timescales comparable to the original one, making photoevaporation a two-timescales process. The bottom left panel shows a disc evolving due to MHD winds (iii): the absolute value of the surface density drops faster than in the viscous model, because of the increase of $\alpha_{\mathrm{DW}}$ in time. Furthermore, as angular momentum is removed from the wind (together with material), the disc does not spread but rather shrinks in time, as expected from the theoretical prediction \citep{Tabone+2022MHDTheory}. Finally, the bottom right panel shows the evolution of a disc undergoing external photoevaporation combined with viscosity (iv): the latter dominates at the earliest stages, producing the characteristic features like viscous spreading, while the effect of external photoevaporation is visible at later ages as a truncation of the disc that also halts its spreading. In this case, the disc truncation and the outside-in depletion of disc material is the consequence of the photo-dissociation of gas molecules due to the FUV radiation emitted by massive stars and experienced by the disc. The efficiency of this process depends primarily on the stellar mass and the FUV flux experienced: given a fixed FUV flux, a disc around a lower mass star will lose its material to external winds more easily compared to a disc around a higher mass star, because of the higher gravitational bond of the system. For the same reason, more extended and less massive discs are more prone to external truncation.

\begin{figure*}
    \centering
    \includegraphics[width = \textwidth]{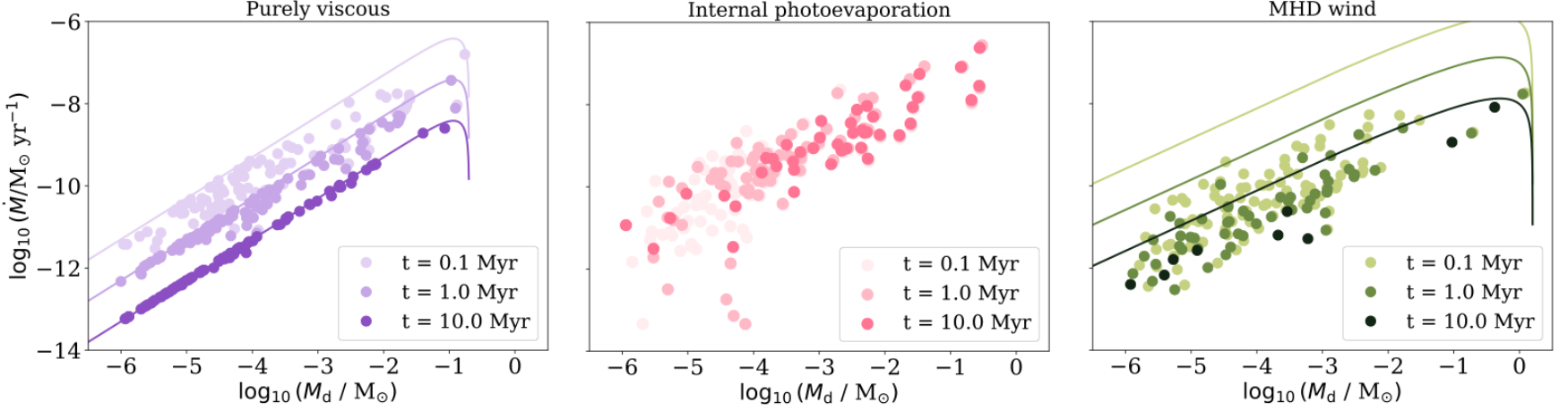}
    \caption{Isochrones at 0.1, 1, and 10 Myr for disc populations undergoing viscous, viscous+internal photoevaporation and wind-driven evolution (left, centre, and right panel respectively), with the same parameters as Figure \ref{fig:validation_sigma}. Each dot represents a disc, while the solid lines show the theoretical isochrones at the corresponding age, where available.}
    \label{fig:validation_isochrones}   
\end{figure*}

Figure \ref{fig:validation_isochrones} shows the isochrones in the $\dot M - M_{\mathrm{d}}$ plane at 0.1, 1 and 10 Myr for the three evolutionary models of viscosity, viscosity and photoevaporation, and MHD winds. Each dot represents a disc in the population, while the solid lines show the analytical prediction (when applicable): in the viscous case, the isochrones read \citep{Lodato2017}

\begin{equation}
    \dot M = \frac{M_{\rm{d}}}{2(2 - \gamma)t} \left[ 1 - \left( \frac{M_{\rm{d}}}{M_0} \right)^{2(2-\gamma)} \right],
    \label{eq:theoretical_isochrone_viscous}
\end{equation}

\noindent while in the MHD wind-driven scenario \citep{Tabone+2022MHDTheory}

\begin{equation}
    \dot M = \frac{1}{\omega (1 + f_{\rm{M}, 0})t} M_{\rm{d}} \left[ \left( \frac{M_{\rm{d}}}{M_0} \right)^{- \omega} - 1 \right].
    \label{eq:theoretical_isochrone_MHD}
\end{equation}

The left panel shows the viscous model, where the discs tend more and more towards the theoretical isochrone during their evolution \citep{Lodato2017}; the central panel includes internal photoevaporation, which has the effect of bending the isochrones once the accretion rate becomes comparable to the photoevaporative mass-loss rate \citep{Somigliana+2020}; finally, the right panel shows an MHD wind-driven population, where the scatter in the $\dot M - M_{\mathrm{d}}$ plane remains significant during the evolution - contrary to the viscous prediction \citep{Somigliana+2023}.

\end{appendix}

\end{document}